\documentclass[aps,reprint,groupedaddress]{revtex4-2}

\usepackage{amsmath}
\usepackage{amssymb}
\usepackage{mathrsfs}

\usepackage{txfonts}
\usepackage{graphicx}

\DeclareMathOperator{\Tr}{\mathop{Tr}}
\DeclareMathOperator{\arcsch}{\mathop{arcsch}}

\begin{document}

\title{Frustration and ordering in Ising chain in an external magnetic field with third-neighbor interactions}

\author{A. V. Zarubin}
\email{Alexander.Zarubin@imp.uran.ru}

\author{F. A. Kassan-Ogly}

\affiliation{M. N. Mikheev Institute of Metal Physics of Ural Branch of Russian Academy of Sciences, S. Kovalevskoy Street 18, 620108 Ekaterinburg, Russia}


\begin{abstract}
In this paper, the frustration properties of the Ising model on a one-dimensional monoatomic equidistant lattice in an external magnetic field are investigated, taking into account the exchange interactions of atomic spins at the sites of the first, second, and third neighbors. Exact analytical expressions for the thermodynamic functions of the system are obtained by the Kramers--Wannier transfer-matrix method. A magnetic phase diagram of the ground state of such a spin system is constructed and studied thoroughly. The points and lines of frustrations of the system depending on the values and signs of exchange interactions and on an external magnetic field are found. The criteria for the occurrence of magnetic frustrations in the presence of competition between the energies of exchange interactions and an external magnetic field are formulated. The peculiar features are investigated and the values of entropy and magnetization of the ground state of this model are obtained in the frustration regime and beyond it. Various types of behavior of entropy, magnetization, and magnetic susceptibility depending on the model parameters are revealed.
\end{abstract}

\maketitle

\section{Introduction}

At the present time, spin systems with magnetic frustrations are being
studied extremely intensively both theoretically and experimentally~\cite{Kassan-Ogly:2010:,Diep:2020,Lacroix:2011,Sadoc:1999,Kudasov:2012:,Vasiliev:2018:}.

The study of frustrated systems makes it possible to understand the
mechanisms of particular magnetic states, such as a spin liquid, spin
ice, and also explain the existence of various incommensurate, helicoidal,
chiral, and other exotic structures (see, for example, Refs.~\cite{Vojta:2018,Balents:2010,Balz:2016,Broholm:2020,Zvyagin:2013:,Lookman:2018,Starykh:2015}).

Magnetic structures with frustrations have been studied since the
second half of the last century, but the phenomenon of magnetic frustrations
was discovered in the mid-seventies of the twentieth century in magnets
exhibiting unusual properties, which was explained by a strong degeneration
of the ground state of the system and the impossibility of magnetic
ordering even at zero temperature. Such magnets by Gerard Toulouse
in 1977 were called \emph{frustrated}~\cite{Toulouse:1977:1,Toulouse:1977:2}.

A crucial point in the study of frustrated systems is the search for
theoretical solutions that allow us to understand the nature of the
occurrence of frustrations in magnetic systems, as well as to adequately
interpret experimental data on magnetic materials containing information
about new phenomena and their unusual properties.

The Ising model is one of the basic models of the theory of magnetism,
for which there is a well-known set of solutions~\cite{Ising:1925,Brush:1967,Niss:2005}
that allow one to describe some spin systems (see, for example, Refs.~\cite{Wolf:2000,Binek:2003}).

In the present paper, we study the frustrating properties of an one-dimensional
Ising model on a monoatomic equidistant lattice in an external magnetic
field, taking into account the exchange interactions of atomic spins
at the sites of the first, second, and third neighbors. Such a model
makes it possible to obtain an exact solution in the thermodynamic
limit, which allows to qualitatively consider the desired characteristics,
including explaining the properties of magnets caused by frustrations
that are not available for description in the framework of a perturbation
theory~\cite{Baxter:1982}.

Our study of the frustration properties of the model is associated
with the investigation of the full magnetic phase diagram of the ground
state, as well as the behavior of the zero-temperature entropy and
magnetization of the system.

Note that this paper is a continuation of the study of the one-dimensional
Ising model, taking into account the exchange interactions of atomic
spins at the sites of the first, second, and third neighbors without
taking into account an external magnetic field, the paper to be referred
to as~\cite{Zarubin:2020}.

\section{Thermodynamic functions of the Ising chain}

We will consider the one-dimensional classical Ising model in an external
magnetic field, taking into account the exchange interactions between
atomic spins at the sites of the first (nearest), second (next-nearest),
and third neighbors, which is given by the Hamiltonian of the form
\begin{equation}
\mathscr{H}=-\sum_{p=1}^{b}\sum_{n=1}^{N-p}J_{p}\sigma_{n}\sigma_{n+p}-\mu_{0}gH\sum_{n=1}^{N}\sigma_{n},\label{eq:H:0}
\end{equation}
where $b$ is the number of exchange interactions of the chain spins
in the model (in this case $b=3$), $J_{1}$ is the parameter of the
exchange interaction between the spins at the nearest neighbor sites
of the linear lattice, $J_{2}$ is the parameter of the exchange interaction
between the spins at the next-nearest neighbor lattice sites, $J_{3}$
is the parameter of the exchange interaction between the spins at
the sites of the third neighbors sites of the lattice, $H$ is the
value of an external uniform magnetic field (directed along the $z$-axis),
$\mu_{0}$ is the Bohr magneton, $g$ is the Land\'{e} $g$-factor, the
symbol $\sigma_{n}$ denotes the $z$-projection of the atomic spin
operator located at the $n$-site and is equal to $\sigma=\pm1(\uparrow,\downarrow)$, 
and $N$ is the number of sites in the spin chain.

In the Kramers--Wannier transfer matrix method \cite{Kramers:1941:1,Baxter:1982}
used and with the imposition of cyclic Born--von K\'{a}rm\'{a}n boundary conditions,
the partition function is equal to
\begin{equation}
Z=\Tr\mathbf{V}^{N},\label{eq:PF}
\end{equation}
where $\mathbf{V}$ is a transfer matrix the elements of which are
independent of the site index~\cite{Baxter:1982} and are specified
by the rule
\begin{multline*}
V_{\sigma^{\prime\prime\prime}\sigma^{\prime\prime\prime\prime}\sigma^{\prime\prime\prime\prime\prime}}^{\sigma\sigma^{\prime}\sigma^{\prime\prime}}=\langle\sigma\sigma^{\prime}\sigma^{\prime\prime}|e^{K_{1}\sigma\sigma^{\prime}+K_{2}\sigma\sigma^{\prime\prime}+K_{3}\sigma\sigma^{\prime\prime\prime}+B\sigma}|\sigma^{\prime\prime\prime}\sigma^{\prime\prime\prime\prime}\sigma^{\prime\prime\prime\prime\prime}\rangle=\\
=e^{K_{1}\sigma\sigma^{\prime}+K_{2}\sigma\sigma^{\prime}+K_{3}\sigma\sigma^{\prime\prime\prime}+B\sigma}\delta_{\sigma^{\prime}\sigma^{\prime\prime\prime}}\delta_{\sigma^{\prime\prime}\sigma^{\prime\prime\prime\prime}}
\end{multline*}
through dimensionless coefficients
\[
K_{1,2,3}=\beta J_{1,2,3},\quad B=\beta\mu_{0}gH,\quad\beta=\frac{1}{k_{\text{B}}T},
\]
and $\delta_{\sigma^{\prime}\sigma^{\prime\prime}}$ is the Kronecker
symbol.

Note that hereinafter such quantities as the Bohr magneton ($\mu_{0}$),
the Land\'{e} $g$-factor ($g$), and the Boltzmann constant ($k_{\text{B}}$)
will be put equal to unity, and the quantities are $T$, $H$, $J_{2}$,
and $J_{3}$ will be measured in $|J_{1}|$ units, as is commonly
accepted in the theory of low-dimensional systems.

The dimension of the square transfer matrix of the one-dimensional
spin model is determined by the expression
\[
d=c^{b},
\]
where $c$ is the number of states at the site ($c=2$ in the classical
Ising model), and $b$ is the number of exchange interactions of chain
spins in the problem ($b=3$). Therefore, in the considered problem,
the dimension of the transfer matrix is equal to
\[
d=2^{3}.
\]

The construction of the transfer matrix was carried out according
to the scheme proposed in \cite{Oguchi:1965}, and described in detail
in \cite{Zarubin:2019:,Zarubin:2020}. Hence, we obtain that the transfer
matrix has the following form\begin{widetext}
\begin{equation}
\mathbf{V}=\left(\begin{array}{cccccccc}
V_{1}V_{2}V_{3}V_{H} & \frac{V_{1}V_{2}V_{H}}{V_{3}} & 0 & 0 & 0 & 0 & 0 & 0\\
0 & 0 & \frac{V_{1}V_{3}V_{H}}{V_{2}} & \frac{V_{1}V_{H}}{V_{2}V_{3}} & 0 & 0 & 0 & 0\\
0 & 0 & 0 & 0 & \frac{V_{2}V_{3}V_{H}}{V_{1}} & \frac{V_{2}V_{H}}{V_{1}V_{3}} & 0 & 0\\
0 & 0 & 0 & 0 & 0 & 0 & \frac{V_{3}V_{H}}{V_{1}V_{2}} & \frac{V_{H}}{V_{1}V_{2}V_{3}}\\
\frac{1}{V_{1}V_{2}V_{3}V_{H}} & \frac{V_{3}}{V_{1}V_{2}V_{H}} & 0 & 0 & 0 & 0 & 0 & 0\\
0 & 0 & \frac{V_{2}}{V_{1}V_{3}V_{H}} & \frac{V_{2}V_{3}}{V_{1}V_{H}} & 0 & 0 & 0 & 0\\
0 & 0 & 0 & 0 & \frac{V_{1}}{V_{2}V_{3}V_{H}} & \frac{V_{1}V_{3}}{V_{2}V_{H}} & 0 & 0\\
0 & 0 & 0 & 0 & 0 & 0 & \frac{V_{1}V_{2}}{V_{3}V_{H}} & \frac{V_{1}V_{2}V_{3}}{V_{H}}
\end{array}\right),\label{eq:N3h:TM}
\end{equation}
\end{widetext}
\[
V_{1}=e^{K_{1}},\quad V_{2}=e^{K_{2}},\quad V_{3}=e^{K_{3}},\quad V_{H}=e^{B}.
\]

The characteristic equation of this matrix is defined as
\begin{multline}
\lambda^{8}+a_{7}\lambda^{7}+a_{6}\lambda^{6}+a_{5}\lambda^{5}+a_{4}\lambda^{4}+a_{3}\lambda^{3}+a_{2}\lambda^{2}+a_{1}\lambda+a_{0}=0,\label{eq:N3h:CP1}
\end{multline}
where the coefficients are
\[
a_{7}=-2e^{K_{1}+K_{2}+K_{3}}\cosh B,
\]
\[
a_{6}=2e^{2K_{2}}\sinh[2(K_{1}+K_{3})],
\]
\[
a_{5}=4e^{-K_{1}+K_{2}+K_{3}}\sinh[2(K_{2}-K_{3})]\cosh B,
\]
\[
a_{4}=-2[\cosh(4K_{2})+\cosh(2B)]+4e^{4K_{3}}\cosh^{2}B,
\]
\[
a_{3}=-8e^{K_{1}-K_{2}+K_{3}}\sinh[2(K_{2}+K_{3})]\sinh(2K_{3})\cosh B,
\]
\[
a_{2}=-8e^{-2K_{2}}\sinh[2(K_{1}-K_{3})]\sinh^{2}(2K_{3}),
\]
\[
a_{1}=-16e^{-K_{1}-K_{2}+K_{3}}\sinh^{3}(2K_{3})\cosh B,
\]
\[
a_{0}=16\sinh^{4}(2K_{3}).
\]

In the transfer matrix method in the thermodynamic limit ($N\to\infty$),
the partition function (\ref{eq:PF}) is defined as
\[
Z=\lambda_{1}^{N},
\]
where $\lambda_{1}$ is the principal (single largest positive real)
eigenvalue of the transfer matrix (\ref{eq:N3h:TM}), which is the
corresponding solution of the equation (\ref{eq:N3h:CP1}). Note that
for the type of matrices under consideration, such a solution always
exists by the Frobenius–Perron theorem \cite{Horn:2013,Domb:1960}.

As a result, all thermodynamic functions of the system, including
the Helmholtz free energy per spin,
\[
F=-\frac{T}{N}\ln Z=-T\ln\lambda_{1},
\]
entropy
\begin{equation}
S=-\frac{\partial F}{\partial T}=\ln\lambda_{1}+\frac{T}{\lambda_{1}}\frac{\partial\lambda_{1}}{\partial T},\label{eq:S0}
\end{equation}
heat capacity
\begin{equation}
C=-T\frac{\partial^{2}F}{\partial T^{2}}=2\frac{T}{\lambda_{1}}\frac{\partial\lambda_{1}}{\partial T}+\frac{T^{2}}{\lambda_{1}}\frac{\partial^{2}\lambda_{1}}{\partial T^{2}}-\frac{T^{2}}{\lambda_{1}^{2}}\left(\frac{\partial\lambda_{1}}{\partial T}\right)^{2},\label{eq:CV}
\end{equation}
magnetization
\begin{equation}
M=-\frac{\partial F}{\partial H}=\frac{T}{\lambda_{1}}\frac{\partial\lambda_{1}}{\partial H},\label{eq:M0}
\end{equation}
and magnetic susceptibility
\begin{equation}
\chi=-\frac{\partial^{2}F}{\partial H^{2}}=\frac{\partial M}{\partial H}=-\frac{T}{\lambda_{1}^{2}}\left(\frac{\partial\lambda_{1}}{\partial H}\right)^{2}+\frac{T}{\lambda_{1}}\frac{\partial^{2}\lambda_{1}}{\partial H^{2}}\label{eq:CHI}
\end{equation}
are defined only in terms of the principal eigenvalue of the transfer
matrix \cite{Baxter:1982,Nolting:2009}.

\section{Magnetic phase diagram of the ground state of the system}

The model contains eight variants for the relationship of the parameters
of the exchange interactions between the spins at the sites of the
first, second, and third neighbors of the chain. These relations are
\begin{equation}
(J_{1}<0,J_{2}>0,J_{3}<0),\quad(J_{1}>0,J_{2}>0,J_{3}>0),\label{eq:N3h:Q:41}
\end{equation}
\begin{equation}
(J_{1}<0,J_{2}>0,J_{3}>0),\quad(J_{1}>0,J_{2}>0,J_{3}<0),\label{eq:N3h:Q:14}
\end{equation}
\begin{equation}
(J_{1}<0,J_{2}<0,J_{3}<0),\quad(J_{1}>0,J_{2}<0,J_{3}>0),\label{eq:N3h:Q:32}
\end{equation}
\begin{equation}
(J_{1}<0,J_{2}<0,J_{3}>0),\quad(J_{1}>0,J_{2}<0,J_{3}<0).\label{eq:N3h:Q:23}
\end{equation}
The first two sets (\ref{eq:N3h:Q:41}) correspond to the aggravated
antiferromagnetic and ferromagnetic types of the exchange interactions.
The last six sets of the parameters (\ref{eq:N3h:Q:14})–(\ref{eq:N3h:Q:23})
define the system with competing exchange interactions between spins.

The presence of an external magnetic field complicates the magnetic
phase diagram of the ground state (MPDGS) of the model, which is determined
by the behavior of the minimum energy of the spin system configurations
at zero temperature, depending on the parameters of the model
\[
E_{0}=\min\{E\},
\]
where the configuration energy itself is the internal energy
\[
U=-T^{2}\frac{\partial}{\partial T}\frac{F}{T}=\frac{T^{2}}{\lambda_{1}}\frac{\partial\lambda_{1}}{\partial T},
\]
per lattice site at zero temperature,
\[
E=\lim_{T\to0}U,
\]
which is explicitly specified by the total energy operator of the
system (\ref{eq:H:0}) and is found from the function
\[
E=\frac{1}{m}\sum_{i=1}^{m}\varepsilon_{i},
\]
\[
\varepsilon_{i}=-H\sigma_{i+b}-\sum_{p=1}^{b}J_{p}\frac{\sigma_{i+b-p}\sigma_{i+b}+\sigma_{i+b}\sigma_{i+b+p}}{2},
\]
where $m$ is the number of sites in the configuration, $b$ is the
number of exchange interactions of the chain spins in the problem
($b=3$), $J_{p}$ is the parameter of the exchange interaction between
spins at neighboring sites of the $p$-level.

Building complete MPDGS depending on the parameters of the model is
an intricate problem and has not been fully carried out. In the papers
\cite{Katsura:1973,Muraoka:1998}, only some aspects of the change
in the MPDGS model in the absence and presence of an external magnetic
field were touched upon.

Thus, only seven types of spin configurations with a minimum energy
are realized in the ground state of the system, depending on the signs
of the parameters of the exchange interactions of the chain spins
and the value of an external magnetic field.

The first type of spin configurations is characterized by ferromagnetic
ordering, which at $H=0$ corresponds to the set
\begin{equation}
C_{\text{F}2}=\left\{ \begin{array}{ccc}
\uparrow & \uparrow & \cdots\\
\downarrow & \downarrow & \cdots
\end{array}\right\} ,\label{eq:C:F2}
\end{equation}
consisting of two sequences (with spin projections along and against
the direction of the $z$-axis) with equal energies
\begin{equation}
E_{\text{F}2}=-(J_{1}+J_{2}+J_{3}),\label{eq:N3h:E:F2}
\end{equation}
and when taking into account an external magnetic field ($H>0$) directed
along the $z$~spin projection, the set of configurations already
consists of one sequence (along the direction of the $z$-axis)
\begin{equation}
C_{\text{F}1}=\left\{ \begin{array}{ccc}
\uparrow & \uparrow & \cdots\end{array}\right\} \label{eq:C:F1}
\end{equation}
with energy
\begin{equation}
E_{\text{F}1}=-(J_{1}+J_{2}+J_{3}+H).\label{eq:N3h:E:F1}
\end{equation}
For such configurations, we introduce index designations F2 and F1,
used in~\cite{Zarubin:2019:,Zarubin:2020}.

The second type of spin configurations is characterized by antiferromagnetic
ordering (configuration designation A2) with the set
\begin{equation}
C_{\text{A}2}=\left\{ \begin{array}{ccc}
\uparrow & \downarrow & \cdots\\
\downarrow & \uparrow & \cdots
\end{array}\right\} ,\label{eq:C:A2}
\end{equation}
consisting of two sequences (with alternating spin projections along
and against the direction of the $z$-axis) with equal energies
\begin{equation}
E_{\text{A}2}=J_{1}-J_{2}+J_{3}.\label{eq:N3h:E:A2}
\end{equation}
This configuration has the indicated energy both in and without an
external magnetic field.

The third type of spin configurations is characterized by magnetic
ordering with a tripling of the translation period (configuration
designation A3), which at $H=0$ has the following set
\begin{equation}
C_{\text{A}3}=\left\{ \begin{array}{cccc}
\uparrow & \uparrow & \downarrow & \cdots\\
\uparrow & \downarrow & \uparrow & \cdots\\
\downarrow & \uparrow & \uparrow & \cdots\\
\downarrow & \downarrow & \uparrow & \cdots\\
\downarrow & \uparrow & \downarrow & \cdots\\
\uparrow & \downarrow & \downarrow & \cdots
\end{array}\right\} ,\label{eq:C:A3}
\end{equation}
consisting of six configurations with equal energies
\begin{equation}
E_{\text{A}3}=\frac{J_{1}+J_{2}-3J_{3}}{3},\label{eq:N3h:E:A3}
\end{equation}
and for $H>0$, the set of sequences is halved and consists of
\begin{equation}
C_{\text{A}31}=\left\{ \begin{array}{cccc}
\uparrow & \uparrow & \downarrow & \cdots\\
\uparrow & \downarrow & \uparrow & \cdots\\
\downarrow & \uparrow & \uparrow & \cdots
\end{array}\right\} \label{eq:C:A31}
\end{equation}
(configuration designation A31). Such a set consists of three configurations
with equal energies
\begin{equation}
E_{\text{A}31}=\frac{J_{1}+J_{2}-3J_{3}-H}{3}.\label{eq:N3h:E:A31}
\end{equation}

The fourth type of configurations is determined by magnetic ordering
with a quadrupling of the period of translations (configuration designation
A4),
\begin{equation}
C_{\text{A}4}=\left\{ \begin{array}{ccccc}
\uparrow & \uparrow & \downarrow & \downarrow & \cdots\\
\uparrow & \downarrow & \downarrow & \uparrow & \cdots\\
\downarrow & \uparrow & \uparrow & \downarrow & \cdots\\
\downarrow & \downarrow & \uparrow & \uparrow & \cdots
\end{array}\right\} ,\label{eq:C:A4}
\end{equation}
which consists of four configurations with equal energies
\begin{equation}
E_{\text{A}4}=J_{2},\label{eq:N3h:E:A4}
\end{equation}
regardless of the presence or absence of an external magnetic field.

The fifth type of configurations is determined by magnetic ordering
with a quadrupling of the period of translations (configuration designation
A41),
\begin{equation}
C_{\text{A}41}=\left\{ \begin{array}{ccccc}
\uparrow & \uparrow & \uparrow & \downarrow & \cdots\\
\uparrow & \uparrow & \downarrow & \uparrow & \cdots\\
\uparrow & \downarrow & \uparrow & \uparrow & \cdots\\
\downarrow & \uparrow & \uparrow & \uparrow & \cdots
\end{array}\right\} ,\label{eq:C:A41}
\end{equation}
which consists of four configurations with equal energies
\begin{equation}
E_{\text{A}41}=-\frac{H}{2}.\label{eq:N3h:E:A41}
\end{equation}

The sixth type of configurations is characterized by magnetic ordering
with quintupling of the period of translations (configuration designation
A5),
\begin{equation}
C_{\text{A}5}=\left\{ \begin{array}{cccccc}
\uparrow & \uparrow & \uparrow & \downarrow & \downarrow & \cdots\\
\uparrow & \uparrow & \downarrow & \downarrow & \uparrow & \cdots\\
\uparrow & \downarrow & \downarrow & \uparrow & \uparrow & \cdots\\
\downarrow & \downarrow & \uparrow & \uparrow & \uparrow & \cdots
\end{array}\right\} ,\label{eq:C:A52}
\end{equation}
which consists of five configurations with equal energies
\begin{equation}
E_{\text{A}5}=-\frac{J_{1}-3(J_{2}+J_{3})+H}{5}.\label{eq:N3h:E:A52}
\end{equation}

The seventh type is characterized by magnetic ordering with a sextuple
period of translations (configuration designation A6),
\begin{equation}
C_{\text{A}6}=\left\{ \begin{array}{ccccccc}
\uparrow & \uparrow & \uparrow & \downarrow & \downarrow & \downarrow & \cdots\\
\uparrow & \uparrow & \downarrow & \downarrow & \downarrow & \uparrow & \cdots\\
\uparrow & \downarrow & \downarrow & \downarrow & \uparrow & \uparrow & \cdots\\
\downarrow & \downarrow & \downarrow & \uparrow & \uparrow & \uparrow & \cdots\\
\downarrow & \downarrow & \uparrow & \uparrow & \uparrow & \downarrow & \cdots\\
\downarrow & \uparrow & \uparrow & \uparrow & \downarrow & \downarrow & \cdots
\end{array}\right\} ,\label{eq:C:A6}
\end{equation}
which consists of six configurations with equal energies
\begin{equation}
E_{\text{A}6}=-\frac{J_{1}-J_{2}-3J_{3}}{3}.\label{eq:N3h:E:A6}
\end{equation}

From this we obtain that in the absence of an external magnetic field,
spin configurations that survive in the ground state are $C_{\text{F}2}$,
$C_{\text{A}2}$, $C_{\text{A}3}$, $C_{\text{A}4}$, and $C_{\text{A}6}$.
On the other hand, in an external magnetic field, the following spin
configurations that survive in the ground state are $C_{\text{F}1}$,
$C_{\text{A}2}$, $C_{\text{A}31}$, $C_{\text{A}4}$, $C_{\text{A}41}$,
$C_{\text{A}5}$, and $C_{\text{A}6}$.

Recall that the ground state configurations listed above correspond
to the following designations $\langle\infty\rangle$ ($C_{\text{F}2}$/$C_{\text{F}1}$),
$\langle1\rangle$ ($C_{\text{A}2}$), $\langle12\rangle$ ($C_{\text{A}3}$/$C_{\text{A}31}$),
$\langle2\rangle$ ($C_{\text{A}4}$), $\langle13\rangle$ ($C_{\text{A}41}$),
$\langle23\rangle$ ($C_{\text{A}5}$), $\langle3\rangle$ ($C_{\text{A}6}$),
introduced in the papers \cite{Fisher:1980,Fisher:1981} and widely
used in the ANNNI model \cite{Selke:1985,Selke:1988,Yeomans:1988}.

Other types of magnetic ordering, that is, spin configurations with
septupling or higher increase in the translation period, do not have
a minimum ground state energy at any ratios of the exchange parameters
of the system.

Thus, the considered spin configurations have the corresponding minimum
energies in the ground state (at $T=0$) in the following ranges of
the model parameters
\begin{equation}
E_{0}=\begin{cases}
E_{\text{F}1}, & P_{\text{F}1},\\
E_{\text{A}2}, & P_{\text{A}2},\\
E_{\text{A}31},\quad & P_{\text{A}31},\\
E_{\text{A}4}, & P_{\text{A}4},\\
E_{\text{A}41}, & P_{\text{A}41},\\
E_{\text{A}5}, & P_{\text{A}5},\\
E_{\text{A}6}, & P_{\text{A}6},
\end{cases}\label{eq:E:0:E}
\end{equation}
where the domains of existence of configurations are limited by the
following conditions
\begin{multline*}
P_{\text{F}1}=\{H\geqslant-2(J_{1}+J_{2})\land H\geqslant-2(J_{1}+J_{3})\\
\land H\geqslant-J_{1}+2(J_{2}+J_{3})\land H\geqslant-(J_{1}+2J_{2}+J_{3})\\
\land H\geqslant-2(J_{1}+J_{2}+J_{3})\land3H\geqslant-2(J_{1}+2J_{2}+3J_{3})\},
\end{multline*}
\begin{multline*}
P_{\text{A}2}=\{J_{1}-J_{2}\leqslant0\land J_{1}-2J_{2}+J_{3}\leqslant0\\
\land H\leqslant-2(J_{1}+J_{3})\land H\leqslant-2(J_{1}-J_{2}+J_{3})\\
\land H\leqslant-2(J_{1}-2J_{2}+3J_{3})\land H\leqslant-2(3J_{1}-4J_{2}+J_{3})\},
\end{multline*}
\begin{multline*}
P_{\text{A}31}=\{H\leqslant-2(J_{1}+J_{2})\land H\geqslant J_{1}-2J_{2}-3J_{3}\\
\land H\leqslant-2(J_{1}+J_{2}-3J_{3})\land H\geqslant-2(J_{1}-2J_{2}+3J_{3})\},
\end{multline*}
\begin{multline*}
P_{\text{A}4}=\{J_{1}-2J_{2}+J_{3}\geqslant0\land J_{1}+2J_{2}-3J_{3}\leqslant0\\
\land H\leqslant-2J_{2}\land H\leqslant J_{1}-2J_{2}-3J_{3}\\
\land H\leqslant-(J_{1}+2J_{2}-3J_{3})\land H\leqslant-(J_{1}+2J_{2}+J_{3})\},
\end{multline*}
\begin{multline*}
P_{\text{A}41}=\{H\geqslant-2J_{2}\land H\leqslant-2(J_{1}+J_{2}+J_{3})\\
\land H\geqslant-2(J_{1}-J_{2}+J_{3})\land H\geqslant-2(J_{1}+J_{2}-3J_{3})\\
\land3H\geqslant2(J_{1}-3(J_{2}+J_{3}))\},
\end{multline*}
\begin{multline*}
P_{\text{A}5}=\{H\leqslant-(J_{1}+2J_{2}+2J_{3})\land H\geqslant-(J_{1}+2J_{2}-3J_{3})\\
\land H\geqslant-2(3J_{1}-4J_{2}+J_{3})\\
\land3H\geqslant2(J_{1}+2J_{2}-3J_{3})\land3H\leqslant2(J_{1}-3J_{2}-3J_{3})\},
\end{multline*}
\begin{multline*}
P_{\text{A}6}=\{J_{1}-J_{2}\geqslant0\land J_{1}+2J_{2}-3J_{3}\geqslant0\\
\land3H\leqslant2(J_{1}+2J_{2}-3J_{3})\land3H\leqslant-2(J_{1}+2J_{2}+3J_{3})\}.
\end{multline*}

\begin{figure*}
\centering\includegraphics{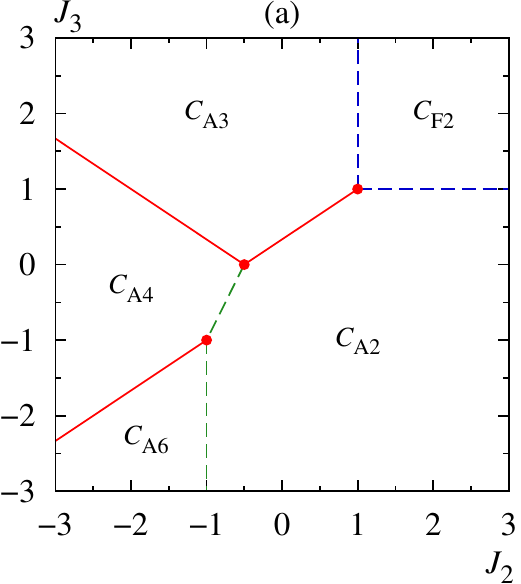}\includegraphics{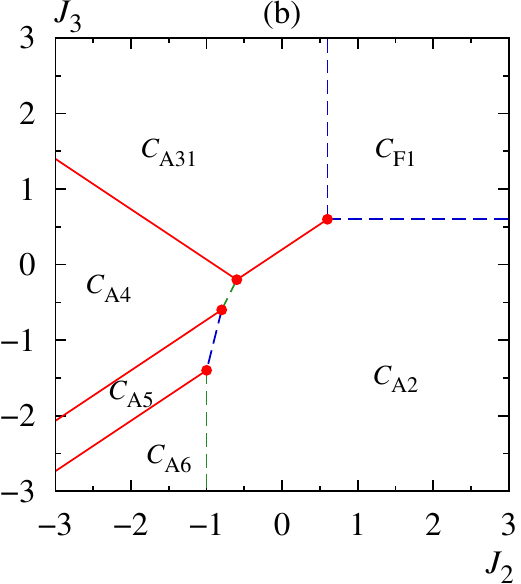}\includegraphics{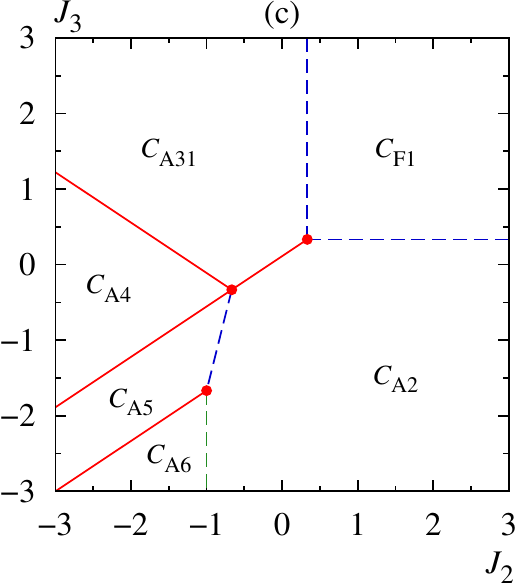}\\
\centering\includegraphics{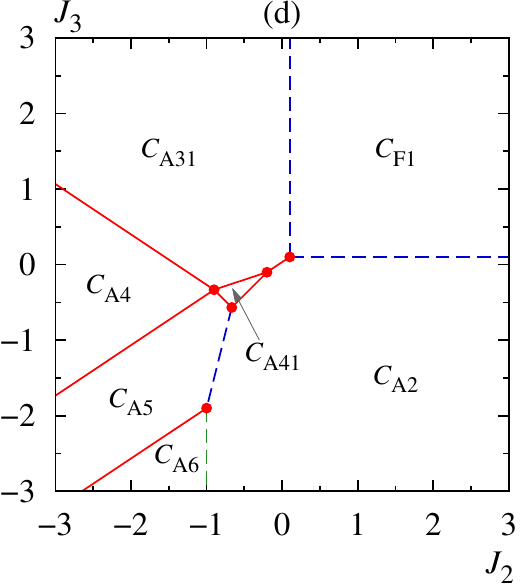}\includegraphics{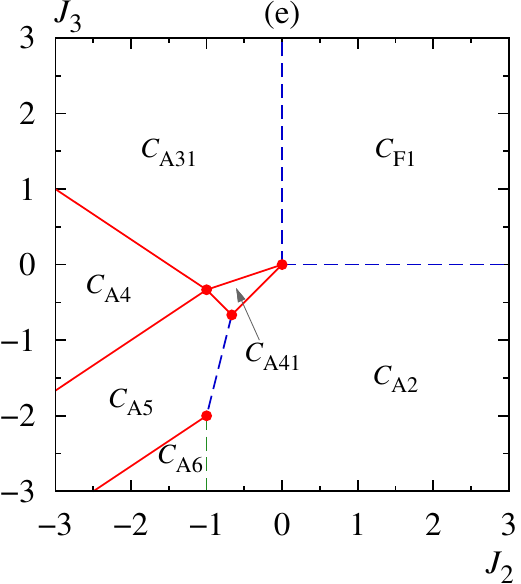}\includegraphics{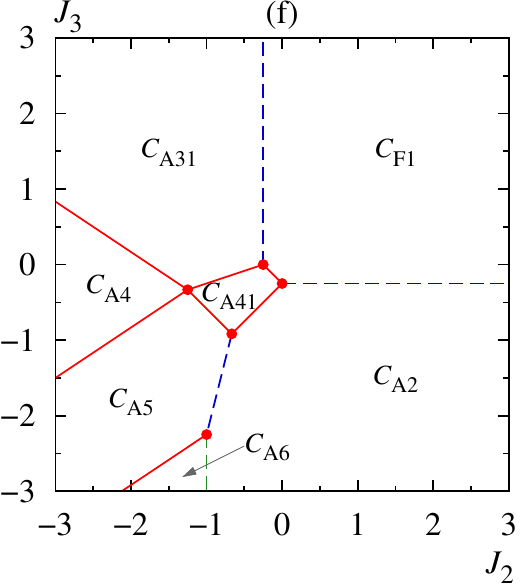}
\caption{MPDGS of the Ising chain in an external magnetic field, taking into
account the exchange interaction of spins at the sites of the first,
second, and third neighbors with antiferromagnetic ($J_{1}=-1$) exchange
interaction of the nearest neighbors, at the values of an external
magnetic field $H=0$~(a), $H=4/5$~(b), $H=4/3$~(c), $H=9/5$~(d),
$H=2$~(e), and $H=5/2$~(f)}
\label{fig:N3h:PD:J3:J2:m}
\end{figure*}

\begin{figure*}
\centering\includegraphics{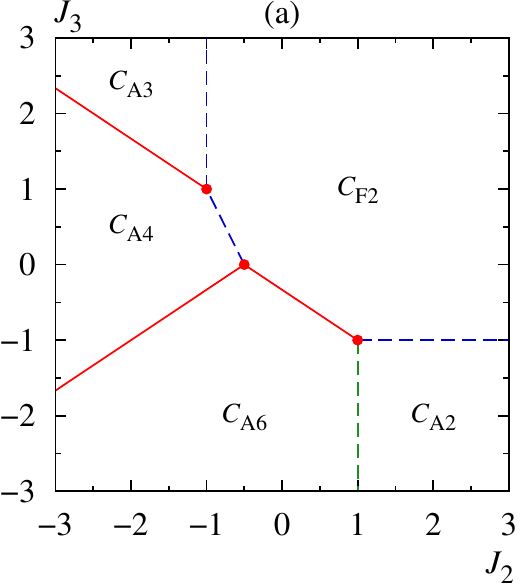}\includegraphics{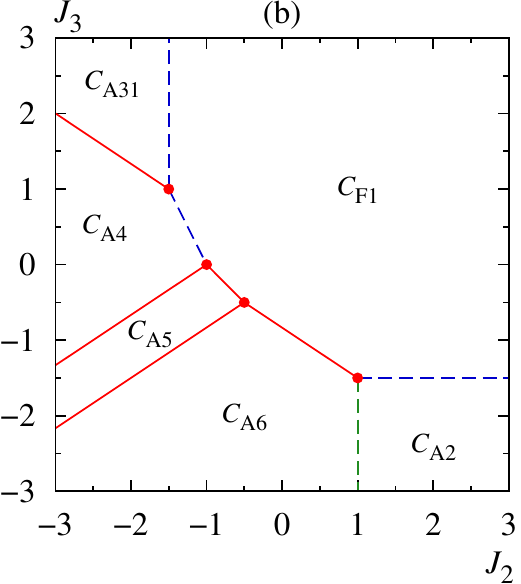}\includegraphics{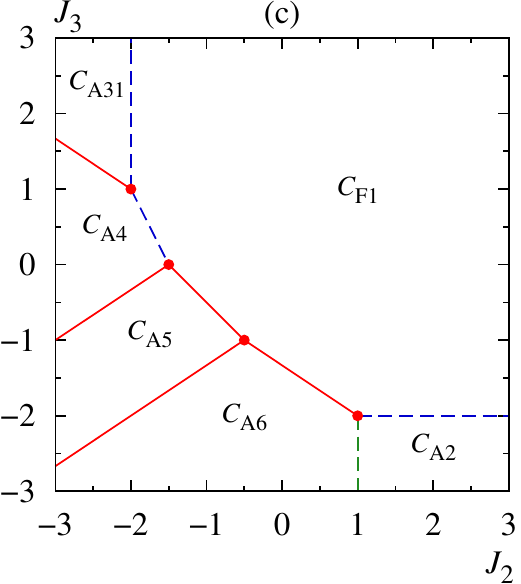}
\caption{MPDGS of the Ising chain in an external magnetic field, taking into
account the exchange interaction of spins at the sites of the first,
second and third neighbors with ferromagnetic ($J_{1}=+1$) exchange
interaction of the nearest neighbors, at the values of an external
magnetic field $H=0$~(a), $H=1$~(b), and $H=2$~(c)}
\label{fig:N3h:PD:J3:J2:p}
\end{figure*}

\begin{figure*}
\centering\includegraphics{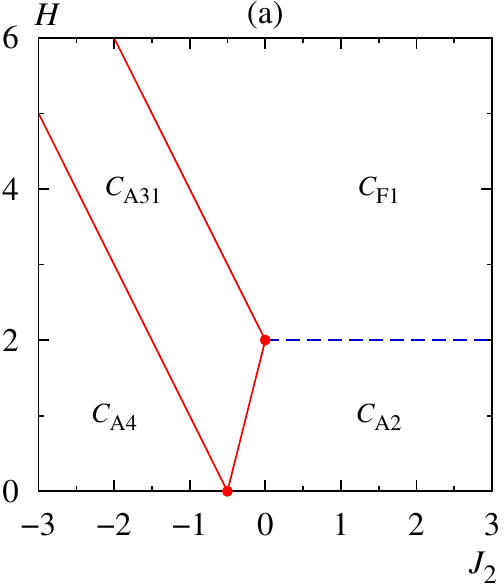}\includegraphics{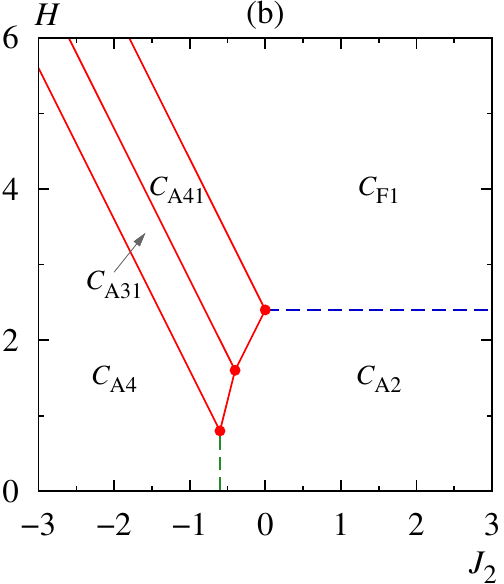}\includegraphics{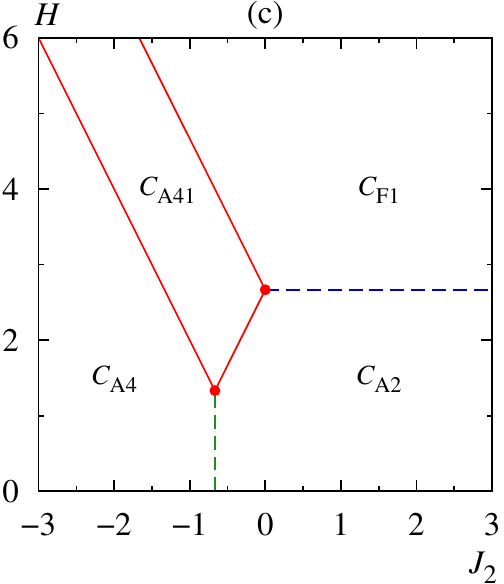}\\
\centering\includegraphics{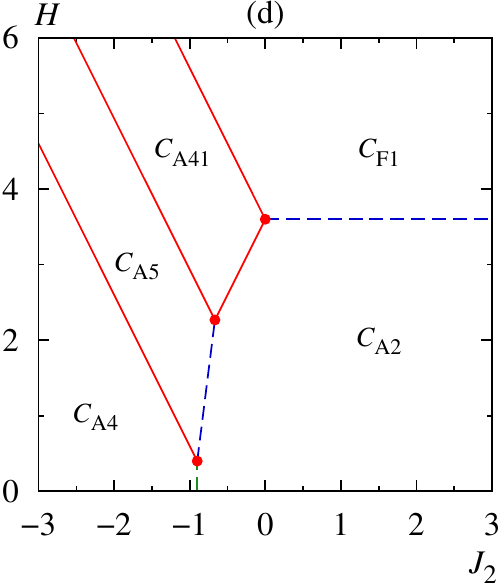}\includegraphics{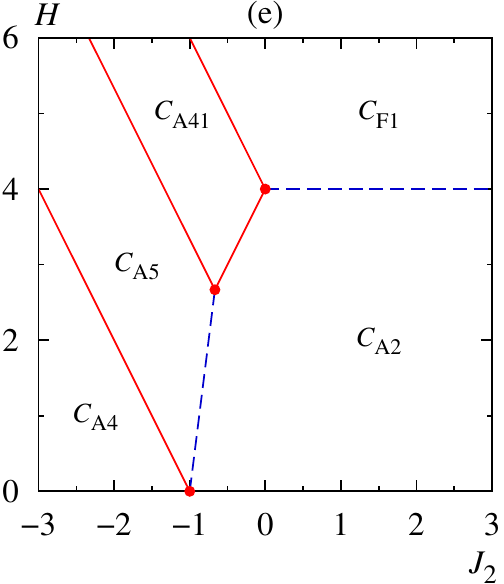}\includegraphics{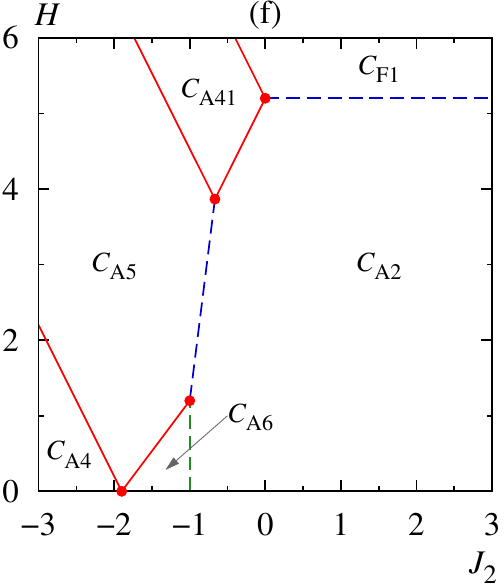}
\caption{MPDGS of the Ising chain in an external magnetic field, taking into
account the exchange interaction of spins at the sites of the first,
second and third neighbors with antiferromagnetic ($J_{1}=-1$) exchange
interaction of nearest neighbors, where $J_{3}=0$~(a), $J_{3}=-1/5$~(b),
$J_{3}=-1/3$~(c), $J_{3}=-4/5$~(d), $J_{3}=-1$~(e), and $J_{3}=-8/5$~(f)}
\label{fig:N3h:PD:H:J2:mm}
\end{figure*}

\begin{figure*}
\centering\includegraphics{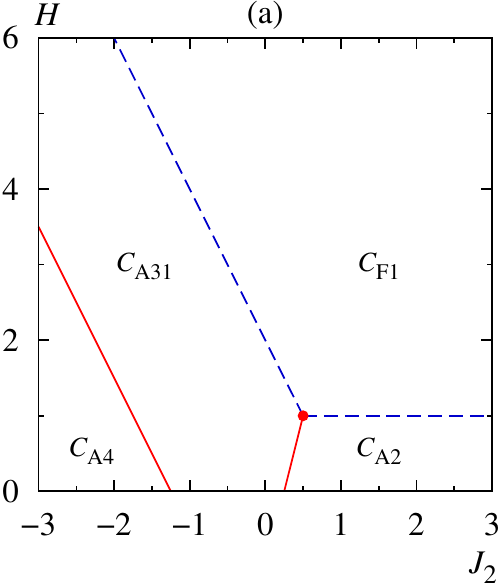}\includegraphics{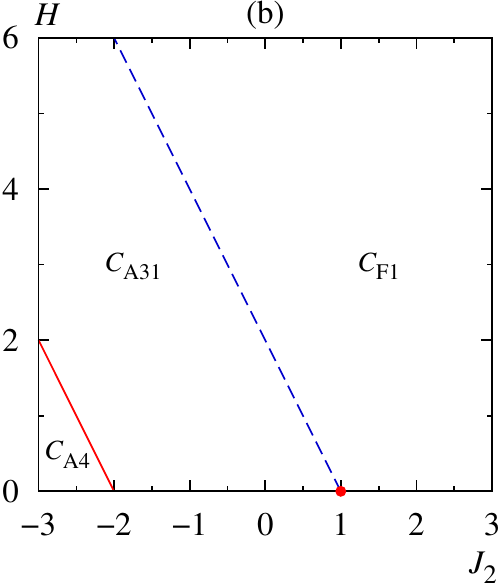}\includegraphics{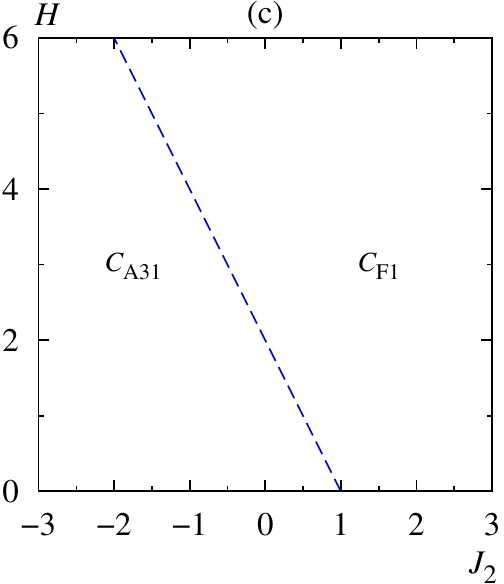}
\caption{MPDGS of the Ising chain in an external magnetic field, taking into
account the exchange interaction of spins at the sites of the first,
second and third neighbors with antiferromagnetic ($J_{1}=-1$) exchange
interaction of the nearest neighbors, where $J_{3}=+1/2$~(a), $J_{3}=+1$~(b),
and $J_{3}=+2$~(c)}
\label{fig:N3h:PD:H:J2:mp}
\end{figure*}

\begin{figure*}
\centering\includegraphics{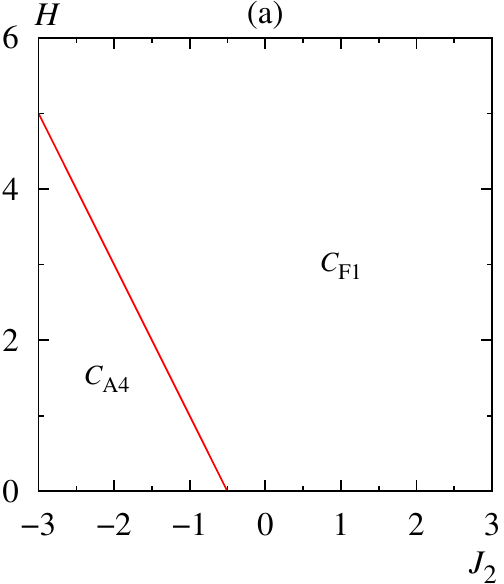}\includegraphics{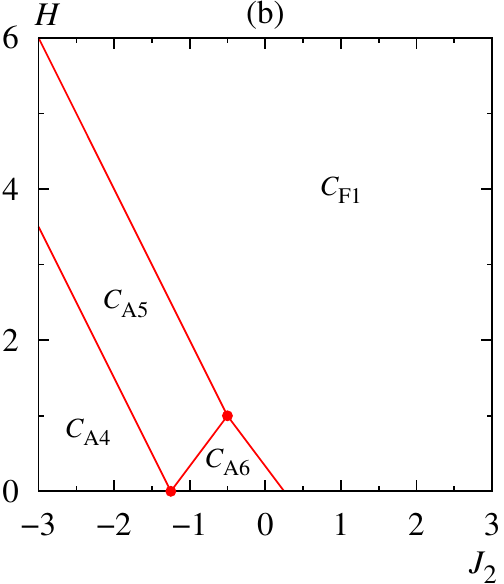}\includegraphics{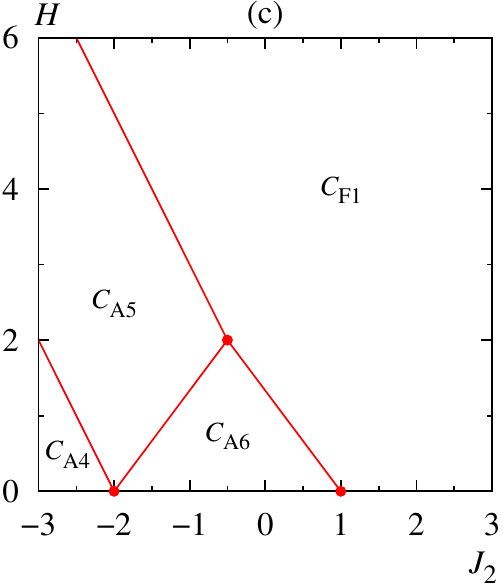}\\
\centering\includegraphics{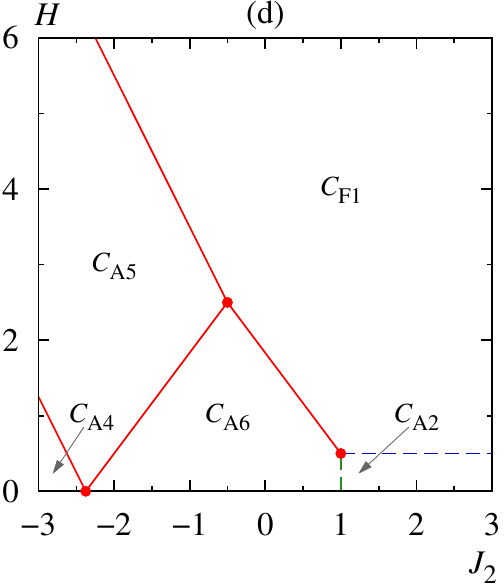}\includegraphics{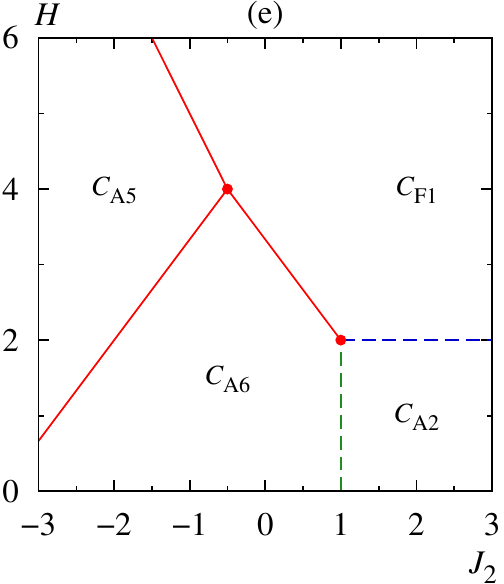}\includegraphics{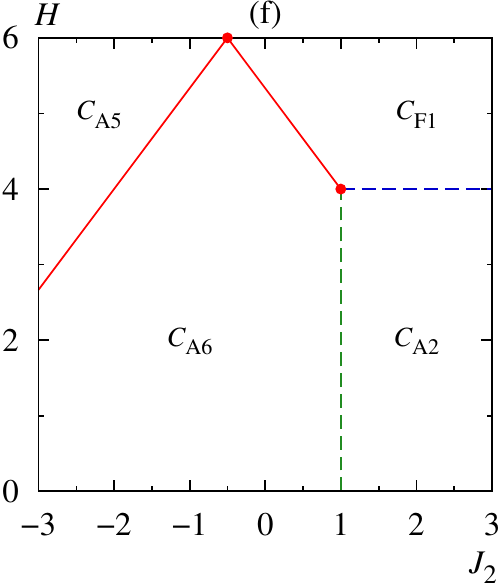}
\caption{MPDGS of the Ising chain in an external magnetic field, taking into
account the exchange interaction of spins at the sites of the first,
second and third neighbors with ferromagnetic ($J_{1}=+1$) exchange
interaction of the nearest neighbors, where $J_{3}=0$~(a), $J_{3}=-1/2$~(b),
$J_{3}=-1$~(c), $J_{3}=-5/4$~(d), $J_{3}=-2$~(e), and $J_{3}=-3$~(f)}
\label{fig:N3h:PD:H:J2:pm}
\end{figure*}

\begin{figure*}
\centering\includegraphics{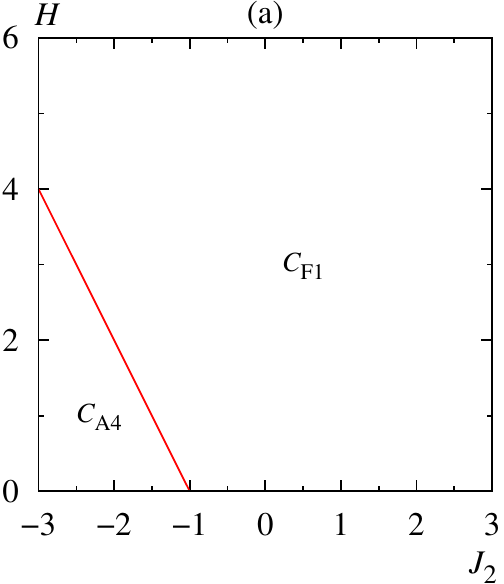}\includegraphics{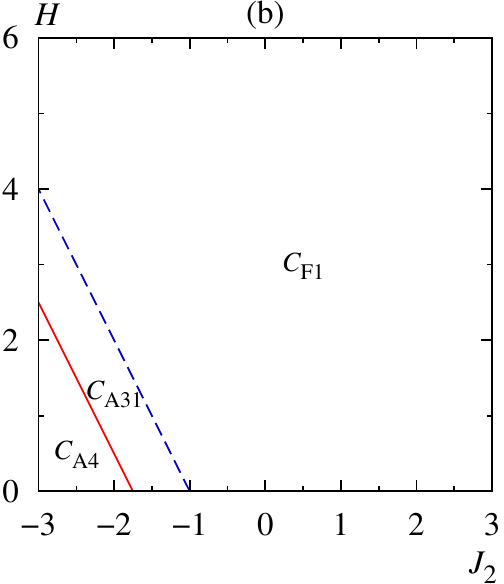}\includegraphics{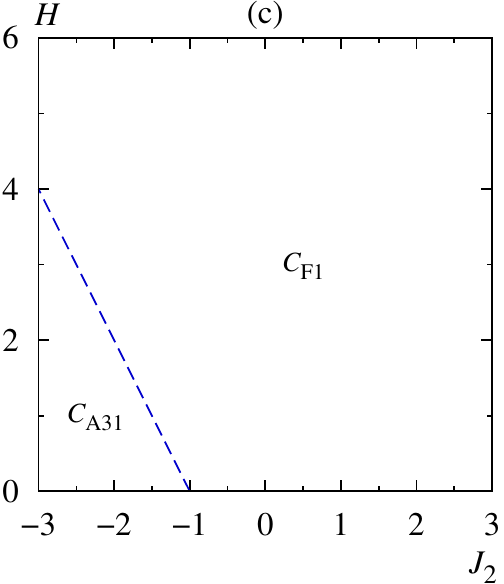}
\caption{MPDGS of the Ising chain in an external magnetic field, taking into
account the exchange interaction of spins at the sites of the first,
second and third neighbors with ferromagnetic ($J_{1}=+1$) exchange
interaction of the nearest neighbors, where $J_{3}=+1$~(a), $J_{3}=+3/2$~(b),
and $J_{3}=+3$~(c)}
\label{fig:N3h:PD:H:J2:pp}
\end{figure*}

\begin{figure*}
\centering\includegraphics{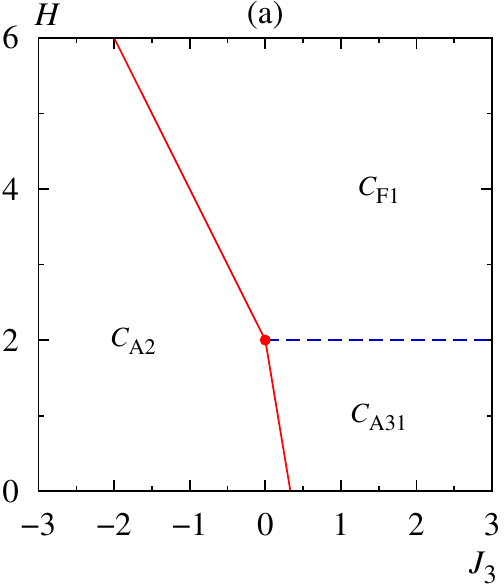}\includegraphics{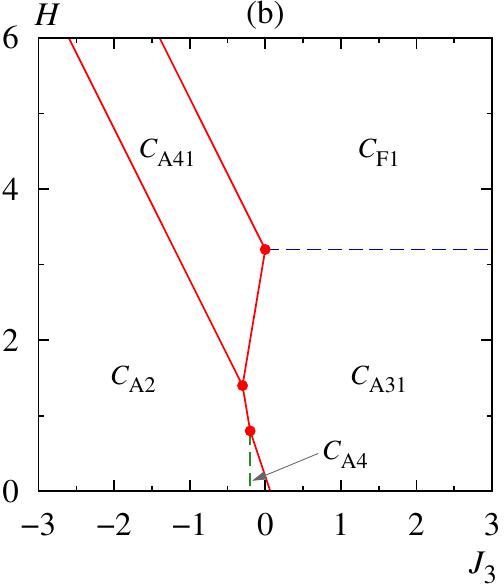}\includegraphics{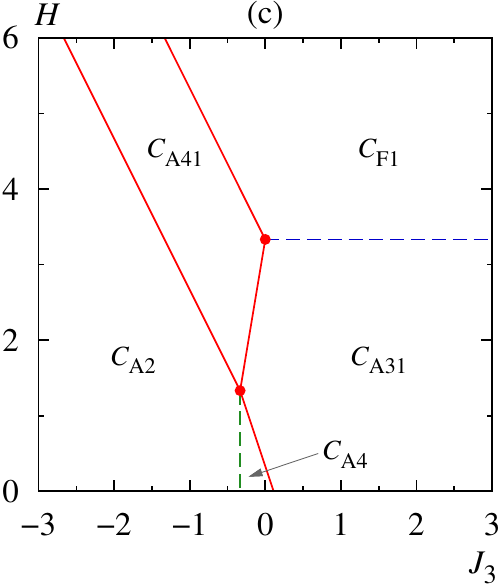}\\
\centering\includegraphics{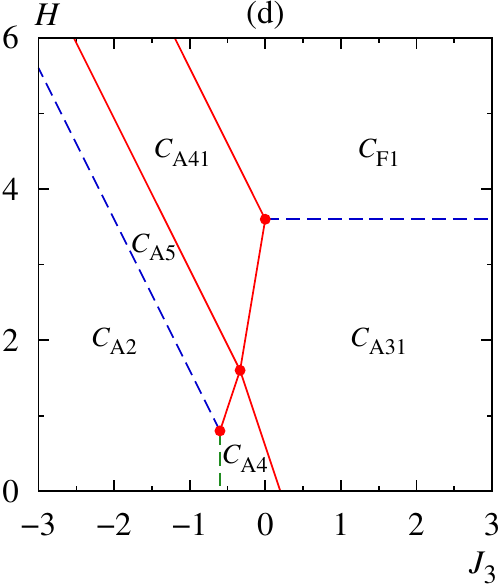}\includegraphics{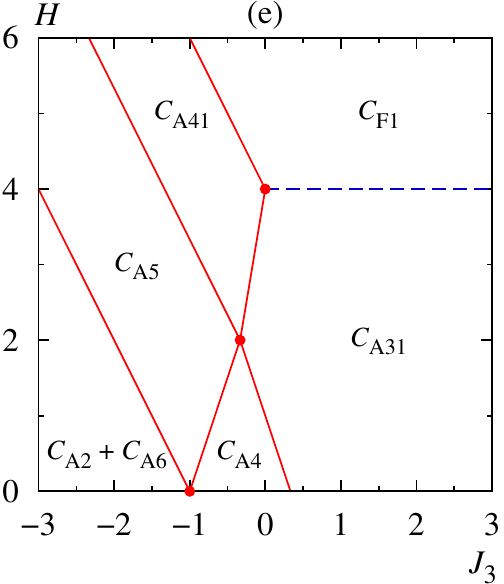}\includegraphics{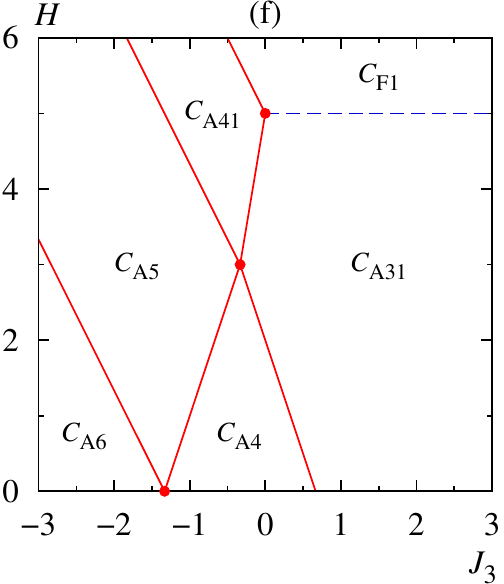}
\caption{MPDGS of the Ising chain in an external magnetic field, taking into
account the exchange interaction of spins at the sites of the first,
second and third neighbors with antiferromagnetic ($J_{1}=-1$) exchange
interaction of nearest neighbors, where $J_{2}=0$~(a), $J_{2}=-3/5$~(b),
$J_{2}=-2/3$~(c), $J_{2}=-4/5$~(d), $J_{2}=-1$~(e), and $J_{2}=-3/2$~(f)}
\label{fig:N3h:PD:H:J3:mm}
\end{figure*}

\begin{figure*}
\centering\includegraphics{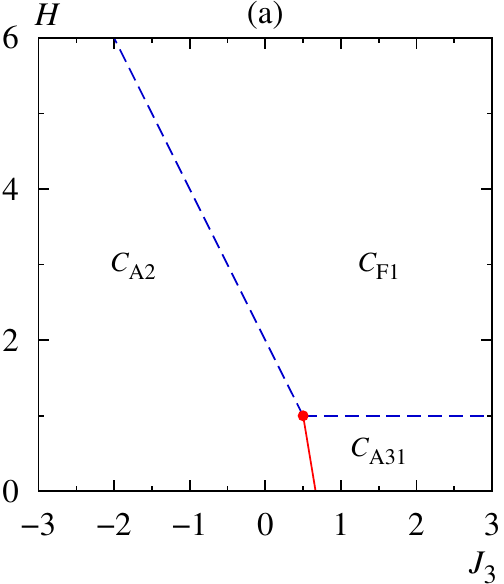}\includegraphics{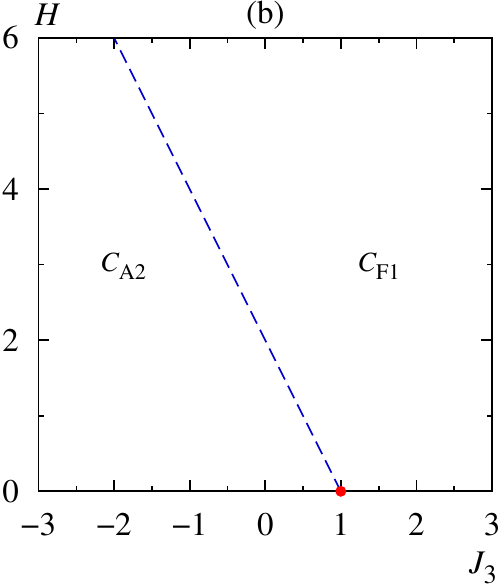}\includegraphics{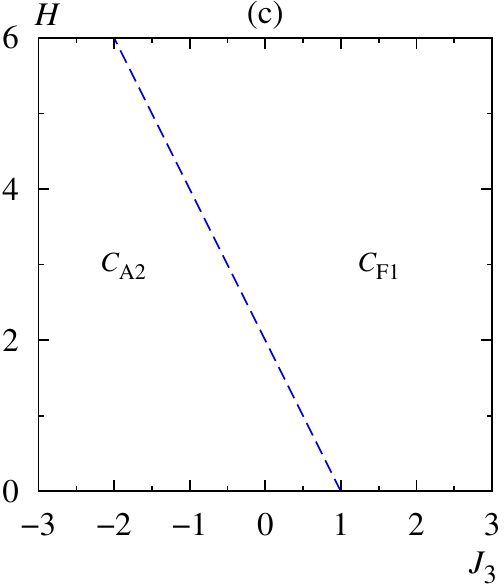}
\caption{MPDGS of the Ising chain in an external magnetic field, taking into
account the exchange interaction of spins at the sites of the first,
second and third neighbors with antiferromagnetic ($J_{1}=-1$) exchange
interaction of the nearest neighbors, where $J_{2}=+1/2$~(a), $J_{2}=+1$~(b),
and $J_{2}=+2$~(c)}
\label{fig:N3h:PD:H:J3:mp}
\end{figure*}

\begin{figure*}
\centering\includegraphics{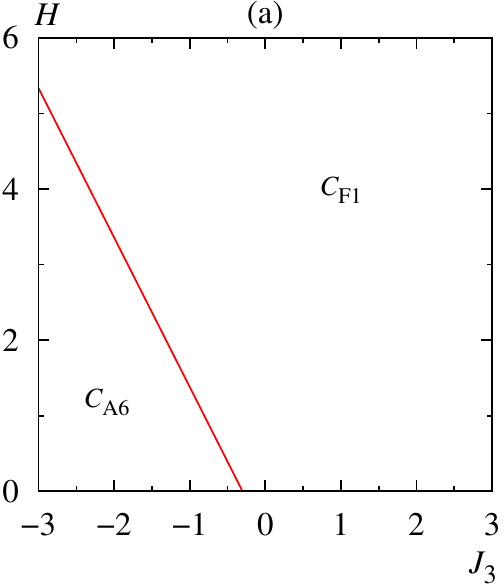}\includegraphics{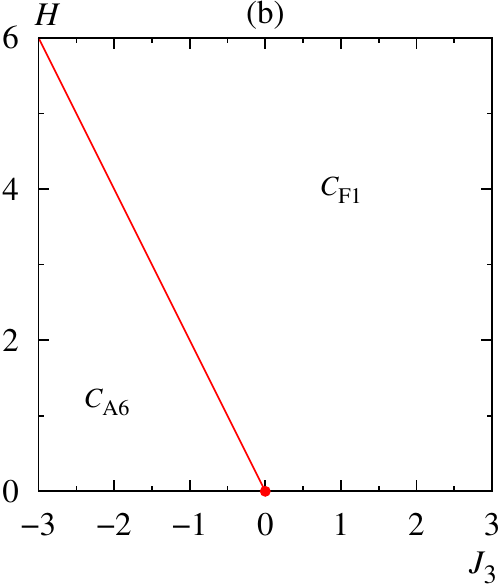}\includegraphics{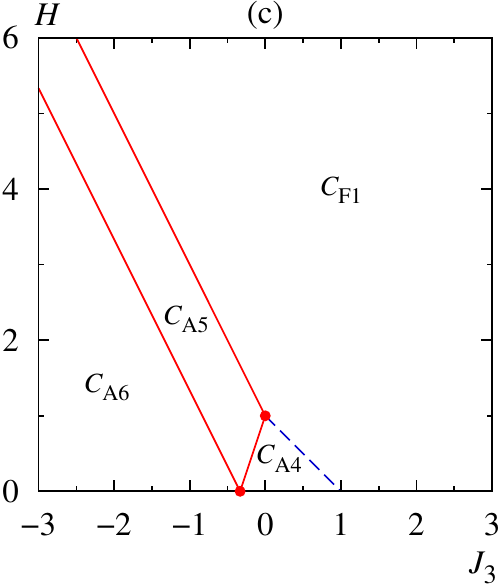}\\
\centering\includegraphics{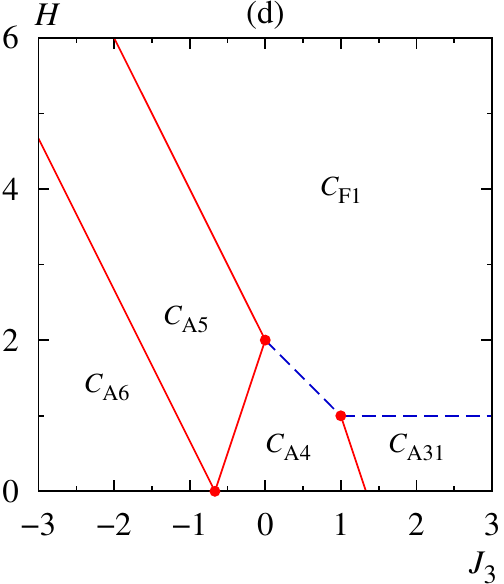}\includegraphics{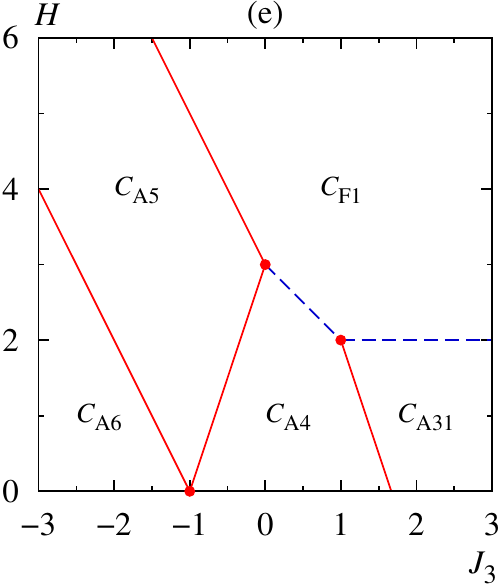}\includegraphics{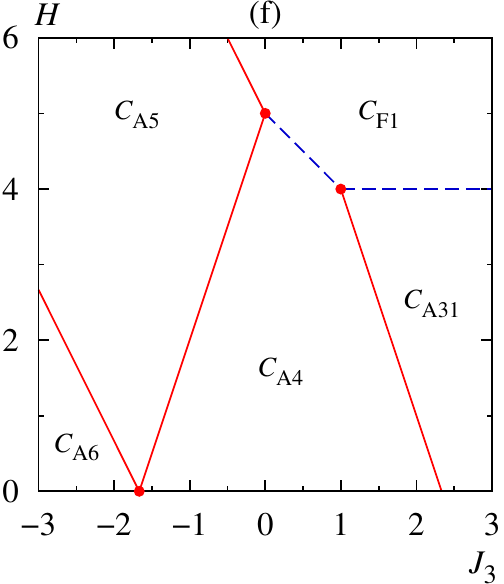}
\caption{MPDGS of the Ising chain in an external magnetic field, taking into
account the exchange interaction of spins at the sites of the first,
second and third neighbors with ferromagnetic ($J_{1}=+1$) exchange
interaction of the nearest neighbors, where $J_{2}=0$~(a), $J_{2}=-1/2$~(b),
$J_{2}=-1$~(c), $J_{2}=-3/2$~(d), $J_{2}=-2$~(e), and $J_{2}=-3$~(f)}
\label{fig:N3h:PD:H:J3:pm}
\end{figure*}

\begin{figure*}
\centering\includegraphics{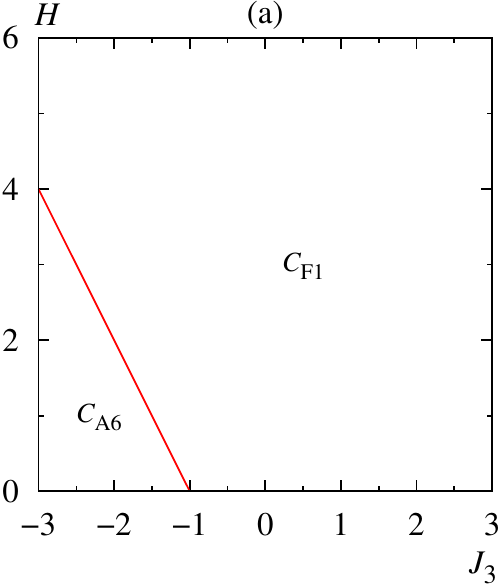}\includegraphics{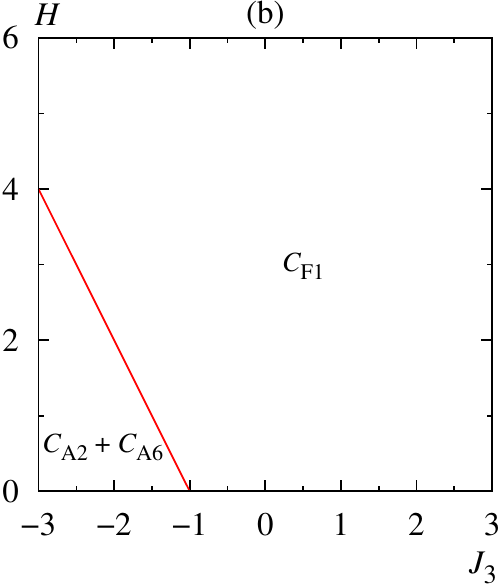}\includegraphics{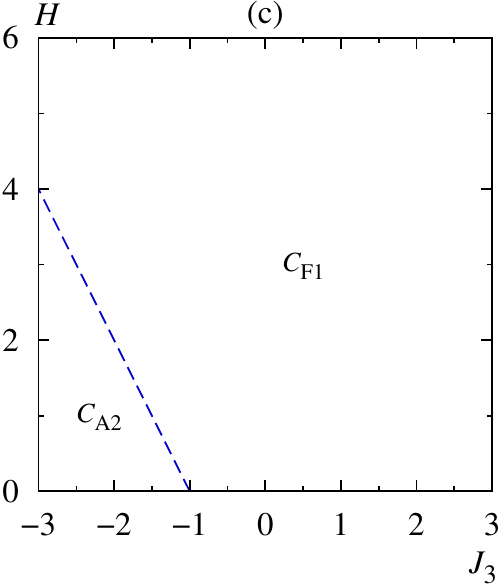}
\caption{MPDGS of the Ising chain in an external magnetic field, taking into
account the exchange interaction of spins at the sites of the first,
second and third neighbors with ferromagnetic ($J_{1}=+1$) exchange
interaction of the nearest neighbors, where $J_{2}=+1/2$~(a), $J_{2}=+1$~(b),
and $J_{2}=+3/2$~(c)}
\label{fig:N3h:PD:H:J3:pp}
\end{figure*}

From the expression for the minimum energy of the ground state (\ref{eq:E:0:E}),
it is easy to obtain all ratios of the parameters of the considered
model, under which the structure of the magnetic ordering of the spin
configurations of the ground state is rearranged with the formation
of the structure of the boundaries of the regions of these configurations
on the MPDGS of the spin system, shown in Figs.~\ref{fig:N3h:PD:J3:J2:m}–\ref{fig:N3h:PD:H:J3:pp}.

As noted earlier, all model parameters are measured in units of $|J_{1}|$,
therefore, for the convenience of constructing phase diagrams, the
quantity $|J_{1}|$ was chosen to be equal to unity. Of course, another
choice of the value of the exchange interaction of the nearest neighbors
$|J_{1}|$ will not fundamentally change of the MPDGS.

The expressions for the minimum energy of spin configurations (\ref{eq:E:0:E})
depend on four model parameters $J_{1}$, $J_{2}$, $J_{3}$, $H$,
so the complete MPDGS is cumbersome.

Note that the presence of competing exchange interactions of spins
at the sites of the first, second, and third neighbors affects the
complexity of the MPDGS itself.

The MPDGS can be plotted as a dependence of the values of the exchange
interactions $J_{3}$ and $J_{2}$ on various values of an external
magnetic field $H$ (Figs.~\ref{fig:N3h:PD:J3:J2:m} and \ref{fig:N3h:PD:J3:J2:p}),
or as a dependence of the values of $H$ and $J_{2}$ on various values
of the exchange interaction $J_{3}$ (Figs.~\ref{fig:N3h:PD:H:J2:mm},
\ref{fig:N3h:PD:H:J2:mp}, \ref{fig:N3h:PD:H:J2:pm}, and \ref{fig:N3h:PD:H:J2:pp}),
and also as a dependence of the values of $H$ and $J_{3}$ on different
values of the exchange interaction $J_{2}$ (Figs.~\ref{fig:N3h:PD:H:J3:mm},
\ref{fig:N3h:PD:H:J3:mp}, \ref{fig:N3h:PD:H:J3:pm}, and \ref{fig:N3h:PD:H:J3:pp}).

Thus, the lines on the MPDGS demonstrate the boundaries of the regions
of spin configurations, at which a qualitative change in the structure
of the magnetic ordering of the ground state of the spin system occurs.

This magnetic phase diagram is complex; it demonstrates not only the
existence of boundaries between two regions of spin configurations,
but also the existence of the intersections of such lines (points)
that delimit three or more regions of spin configurations. (Note that
this situation was considered in detail in \cite{Zarubin:2020,Zarubin:2019:,Zarubin:2019:E}.)

The dashed lines on the MPDGS indicate the boundaries at which the
ground state ordering is rearranged, and the number of configurations
of the system with the minimum energy is equal to the sum of the configurations
of the regions adjacent to the boundary.

The solid lines on the MPDGS indicate the boundaries at which the
number of configurations of the system with the minimum energy is
greater than the sum of the configurations of the regions of the phase
diagram adjacent to it.

The existence of such set of spin configurations of the system at
zero temperature at the boundaries and at points of the phase space
is associated with the rearrangement of the magnetic structure and
the appearance at a given phase point (in the thermodynamic limit)
of an infinite number of spin configurations, including those without
any translational invariance.

Crossing the boundaries of spin configurations on the MPDGS with the
formation of triple, quadruple or with higher multiplicity points
are marked on the plots by round dots.

It should be noted that (in the terminology of the papers~\cite{Fisher:1981,Pokrovskii:1982:,Selke:1985,Barreto:1985,Yeomans:1987,Yeomans:1988})
on the MPDGS, the solid lines described above are called multiphase
lines, and triple, quadruple and points with higher multiplicity are
called multiphase points.

\section{Thermodynamics of the system at zero temperature without an external magnetic field}

Taking into account the exchange interactions between atomic spin
at the sites of the first, second, and third neighbors, and in the
absence of an external magnetic field ($H=0$), the characteristic
equation (\ref{eq:N3h:CP1}) is defined as
\begin{multline}
(\lambda^{4}+b_{3}\lambda^{3}+b_{2}\lambda^{2}+b_{1}\lambda+b_{0})(\lambda^{4}+c_{3}\lambda^{3}+c_{2}\lambda^{2}+c_{1}\lambda+c_{0})=0,\label{eq:N3:CP1}
\end{multline}
where the coefficients are
\[
b_{3}=-2e^{K_{2}}\cosh(K_{1}+K_{3}),\quad c_{3}=-2e^{K_{2}}\sinh(K_{1}+K_{3}),
\]
\[
b_{2}=-c_{2}=2\sinh(2K_{2}),
\]
\[
b_{1}=4e^{-K_{2}}\sinh(2K_{3})\sinh(K_{1}-K_{3}),
\]
\[
c_{1}=-4e^{-K_{2}}\sinh(2K_{3})\cosh(K_{1}-K_{3}),
\]
\[
b_{0}=c_{0}=4\sinh^{2}(2K_{3}).
\]
The principal eigenvalue of the transfer matrix determined from the
equation (\ref{eq:N3:CP1}), is expressed in radicals and has the
following form
\begin{equation}
\lambda_{1}=-\frac{b_{3}}{4}-\Psi+\frac{1}{2}\sqrt{-4\Psi^{2}-2p+\frac{q}{S}},\label{eq:L1}
\end{equation}
\[
p=b_{2}-\frac{3}{8}b_{3}^{2},\quad q=b_{1}-\frac{b_{2}b_{3}}{2}+\frac{b_{3}^{3}}{8},
\]
\[
\Psi=\frac{1}{2}\sqrt{-\frac{2}{3}p+\frac{1}{3}\left(\Theta+\frac{\Delta_{0}}{\Theta}\right)},
\]
\[
\Theta=\sqrt[3]{\frac{\Delta_{1}+\sqrt{\Delta_{1}^{2}-4\Delta_{0}^{3}}}{2}},
\]
\[
\Delta_{0}=12b_{0}-3b_{1}b_{3}+b_{2}^{2},
\]
\[
\Delta_{1}=-72b_{0}b_{2}+27b_{0}b_{3}^{2}+27b_{1}^{2}-9b_{1}b_{2}b_{3}+2b_{2}^{3}.
\]
Using the expression (\ref{eq:L1}) it is possible to calculate all
necessary thermodynamic functions of the system.

We also note that the expression for the minimum energy of spin configurations
(\ref{eq:E:0:E}) makes it possible to obtain the MPDGS of the system,
which is shown in Figs. \ref{fig:N3h:PD:J3:J2:m}a and \ref{fig:N3h:PD:J3:J2:p}a.

In the regions beyond the boundaries of spin configurations $C_{\text{F}2}$,
$C_{\text{A}2}$, $C_{\text{A}3}$, $C_{\text{A}4}$, $C_{\text{A}6}$
on the MPDGS, the zero-temperature (residual) entropy of the system
is always equal to zero,
\begin{equation}
\lim_{T\to0}S\equiv S^{\circ}=0,\label{eq:S0:P}
\end{equation}
and what the system demonstrates in these regions is the equilibrium
state of the system.

At the boundaries of the regions of spin configurations $C_{\text{F}2}$–$C_{\text{A}2}$,
$C_{\text{F}2}$–$C_{\text{A}3}$, $C_{\text{A}2}$–$C_{\text{A}6}$,
$C_{\text{A}2}$– $C_{\text{A}4}$ ($J_{1}<0$), $C_{\text{F}2}$–$C_{\text{A}4}$
($J_{1}>0$), the residual entropy is also equal to zero
\begin{equation}
S^{\circ}=0.\label{eq:S0:L}
\end{equation}

As noted earlier, this is due to the fact that at these boundaries
such a number of configurations of the system with a minimum energy
are formed that is equal to the sum of the configurations of the regions
adjacent to this boundary. Therefore, the residual entropy of the
equilibrium system (according to the Nernst–Planck theorem) is equal
to zero \cite{Sommerfeld:1956,Nolting8:2018}.

Such boundaries with zero residual entropy on the MPDGS are marked
with dashed lines.

Next, we list the cases when the residual entropy is greater than
zero 
\[
S^{\circ}>0
\]
at the junctions of the regions of the spin configurations of the
system on the MPDGS.

Note that this situation is possible and this result does not contradict
the third law of thermodynamics~\cite{Sommerfeld:1956,Nolting8:2018},
and such states of the system in which the entropy of the ground state
is greater than zero are \emph{frustrated} (see the discussion in~\cite{Zarubin:2019:,Zarubin:2020}).

Such boundaries with nonzero residual entropy are marked with solid
lines on the MPDGS, and triple and with higher multiplicity points
are marked with a solid circle. It should also be said that such positions
of the frustration of the system on the MPDGS correspond to multiphase
lines and multiphase points in the terminology of \cite{Barreto:1985,Selke:1985}.

Thus, at the boundaries of the regions of spin configurations $C_{\text{A}3}$–$C_{\text{A}4}$
and $C_{\text{A}4}$–$C_{\text{A}6}$ (Figs. \ref{fig:N3h:PD:J3:J2:m}a
and \ref{fig:N3h:PD:J3:J2:p}a), and at the points $C_{\text{A}2}$–$C_{\text{A}4}$(–$C_{\text{A}5}$)–$C_{\text{A}6}$
($J_{1}<0$, Fig. \ref{fig:N3h:PD:J3:J2:m}a), $C_{\text{F}2}$–$C_{\text{A}3}$–$C_{\text{A}4}$(–$C_{\text{A}5}$)
($J_{1}>0$, Fig. \ref{fig:N3h:PD:J3:J2:p}a), $C_{\text{A}2}$–$C_{\text{A}4}$–$C_{\text{A}5}$(–$C_{\text{A}6}$)
($J_{1}<0$, $H=0$, Fig. \ref{fig:N3h:PD:H:J2:mm}e), $C_{\text{A}4}$–$C_{\text{A}5}$–$C_{\text{A}6}$
($H=0$, Fig. \ref{fig:N3h:PD:H:J2:mm}f, \ref{fig:N3h:PD:H:J2:pm}b),
$C_{\text{F}2}$(–$C_{\text{A}3}$)–$C_{\text{A}4}$ ($J_{1}>0$,
Fig.\ref{fig:N3h:PD:H:J3:pm}c) the residual entropy is
\begin{equation}
S^{\circ}=\ln\left[\frac{1}{3}\left(\vartheta_{1}+\frac{3}{\vartheta_{1}}\right)\right]\approx0.281\,199\,6,\label{eq:N3h:S0:281}
\end{equation}
where
\begin{equation}
\vartheta_{1}=\sqrt[3]{\frac{3^{3}+3\sqrt{-3\mathcal{D}_{1}}}{2}}.\label{eq:N3h:S0:281:T1}
\end{equation}
Note that the sublogarithmic expression in Eq. (\ref{eq:N3h:S0:281})
is the single positive largest real (principal) solution of the equation
\begin{equation}
x^{3}-x-1=0,\label{eq:N3h:S0:281:E0}
\end{equation}
its discriminant is
\[
\mathcal{D}_{1}=2^{2}-3^{3}=-23.
\]

At the boundaries of the regions of spin configurations $C_{\text{A}2}$–$C_{\text{A}3}$
($J_{1}<0$, Fig. \ref{fig:N3h:PD:J3:J2:m}a) and $C_{\text{F}2}$–$C_{\text{A}6}$
($J_{1}>0$, Fig. \ref{fig:N3h:PD:J3:J2:p}a), and also at the points
$C_{\text{F}2}$–$C_{\text{A}2}$–$C_{\text{A}3}$ ($J_{1}<0$, Fig.
\ref{fig:N3h:PD:J3:J2:m}a) and $C_{\text{F}2}$–$C_{\text{A}2}$–$C_{\text{A}6}$
($J_{1}>0$, Fig. \ref{fig:N3h:PD:J3:J2:p}a), $C_{\text{F}2}$(–$C_{\text{A}2}$)–$C_{\text{A}6}$
($J_{1}>0$, $H=0$, Fig. \ref{fig:N3h:PD:H:J2:pm}c, \ref{fig:N3h:PD:H:J3:pp}a),
$C_{\text{F}2}$–$C_{\text{A}2}$+$C_{\text{A}6}$ ($J_{1}>0$, $H=0$,
Fig. \ref{fig:N3h:PD:H:J3:pp}b) the residual entropy is
\begin{equation}
S^{\circ}=\ln\left[\frac{1}{3}\left(1+\vartheta_{2}+\frac{1}{\vartheta_{2}}\right)\right]\approx0.382\,245\,1,\label{eq:N3h:S0:382}
\end{equation}
\begin{equation}
\vartheta_{2}=\sqrt[3]{\frac{2+3^{3}+3\sqrt{-3\mathcal{D}_{2}}}{2}},\label{eq:N3h:S0:382:T1}
\end{equation}
where the sublogarithmic expression is the principal solution of the
equation
\begin{equation}
x^{3}-x^{2}-1=0,\label{eq:N3h:S0:382:E0}
\end{equation}
its discriminant is
\[
\mathcal{D}_{2}=-(2^{2}+3^{3})=-31.
\]

At triple points $C_{\text{A}2}$–$C_{\text{A}3}$–$C_{\text{A}4}$
($J_{1}<0$, Fig. \ref{fig:N3h:PD:J3:J2:m}a) and $C_{\text{F}2}$–$C_{\text{A}4}$–$C_{\text{A}6}$
($J_{1}>0$, Fig. \ref{fig:N3h:PD:J3:J2:p}a), and also at the points
$C_{\text{F}2}$–$C_{\text{A}4}$(–$C_{\text{A}5}$–$C_{\text{A}6}$)
($J_{1}>0$, $H=0$, Fig.~\ref{fig:N3h:PD:H:J2:pm}a) the residual
entropy is equal to the natural logarithm of the golden ratio,
\begin{equation}
S^{\circ}=\ln\frac{1+\sqrt{5}}{2}=\arcsch2\approx0.481\,211\,8,\label{eq:N3h:S0:481}
\end{equation}
where the sublogarithmic expression in (\ref{eq:N3h:S0:481}) is the
principal solution of the equation
\begin{equation}
x^{2}-x-1=0,\label{eq:N3h:S0:481:E0}
\end{equation}
its discriminant is $\mathcal{D}_{3}=5$.

\begin{figure}[ht]
\centering\includegraphics{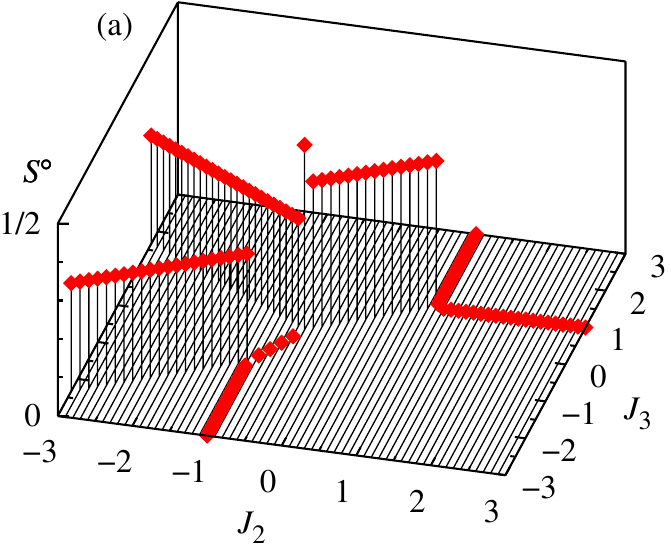}\quad{}\includegraphics{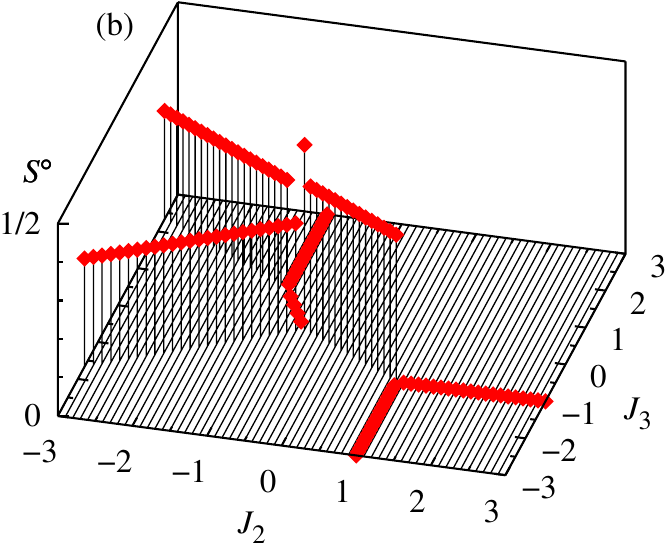}
\caption{Entropy of the ground state of the Ising chain without an external
magnetic field, taking into account the interaction of spins at the
sites of the first, second, and third neighbors with antiferromagnetic
($J_{1}=-1$) (a) and ferromagnetic ($J_{1}=+1$) (b) interactions
of nearest neighbors. Red rhombic dots indicate the values of entropy
at the boundaries of the spin configurations}
\label{fig:N3h:S0:J3:J2}
\end{figure}

An example of the behavior of the residual entropy of the Ising chain
in the absence of an external magnetic field is shown in Fig.~\ref{fig:N3h:S0:J3:J2}.
Comparison of this plot with the magnetic phase diagram in Figs.~\ref{fig:N3h:PD:J3:J2:m}a
and \ref{fig:N3h:PD:J3:J2:p}a gives a complete picture of the behavior
of entropy at zero temperature.

It should also be said that in the case under consideration at zero
temperature ($T=0$) in the absence of an external magnetic field
($H=0$) other considered thermodynamic functions, such as residual
heat capacity (\ref{eq:CV}) and residual magnetization (\ref{eq:M0})
systems are equal to zero
\[
C^{\circ}=0,\quad M^{\circ}=0,
\]
respectively.

In turn, the residual magnetic susceptibility (\ref{eq:CHI}) of the
system beyond the boundaries of spin configurations and at the boundaries
$C_{\text{A}2}$–$C_{\text{A}4}$ and $C_{\text{A}2}$–$C_{\text{A}6}$
on the MPDGS is zero,
\[
\chi^{\circ}=0
\]
(green dashed lines in Figs.~\ref{fig:N3h:PD:J3:J2:m}a and \ref{fig:N3h:PD:J3:J2:p}a).
At the boundaries of spin configurations $C_{\text{F}2}$–$C_{\text{A}2}$,
$C_{\text{F}2}$–$C_{\text{A}3}$, and $C_{\text{F}2}$–$C_{\text{A}4}$
the residual magnetic susceptibility has values in the interval
\[
0<\chi^{\circ}<\infty
\]
(blue dashed lines in Figs.~\ref{fig:N3h:PD:J3:J2:m}a and \ref{fig:N3h:PD:J3:J2:p}a).
In the case of frustrated states, the residual magnetic susceptibility
is equal to infinity,
\[
\chi^{\circ}=\infty
\]
(red solid lines in Figs.~\ref{fig:N3h:PD:J3:J2:m}a and \ref{fig:N3h:PD:J3:J2:p}a).

Note that a detailed study of the temperature dependences of the thermodynamic
functions of this model without taking into account an external magnetic
field was done in \cite{Zarubin:2020}.

At the end of this section, we note that in the absence of exchange
between the spins of the chain ($J_{i}=0$), a quintuple point is
formed on the MPDGS, in which a paramagnetic state is realized, characterized
by the fact that all configurations of the system have the same probability
and have the same zero energy. The entropy of such a state of the
system is equal to the natural logarithm of two,
\begin{equation}
S=\ln2\approx0.693\,147\,2,\label{eq:N3h:S0:0}
\end{equation}
and it is the same (maximum) at any temperature. From this it is clear
that the \emph{Ising paramagnet is an absolutely frustrated system}
(see discussion in \cite{Zarubin:2019:,Zarubin:2020}).

We also recall that in the expression for the Gibbs entropy, the argument
of the natural logarithm is the statistical weight of the system ($W$).
In the case of (\ref{eq:N3h:S0:0}) $W=2$, which determines the number
of possible configurations of the considered model.

\section{Thermodynamics of the system at zero temperature in an external magnetic field}

The inclusion of an external magnetic field exceedingly complicates
the behavior of the system. In addition to entropy and heat capacity,
one can find the magnetization and magnetic susceptibility of the
spin system.

In all possible regions of spin configurations ($C_{\text{F}1}$,
$C_{\text{A}2}$, $C_{\text{A}31}$, $C_{\text{A}4}$, $C_{\text{A}41}$,
$C_{\text{A}5}$, $C_{\text{A}6}$), the residual entropy is always
equal to zero, and the zero-temperature (residual) magnetization has,
among other things, non-zero values.

The residual magnetization of regions with period doubling (antiferromagnetic
ordering), period quadrupling and sextupling is equal to zero
\[
M_{\text{A}2}^{\circ}=0,\quad M_{\text{A}4}^{\circ}=0,\quad M_{\text{A}6}^{\circ}=0,
\]
and more complicated configurations with period tripling, quadrupling,
and quintupling have the following residual magnetizations
\[
M_{\text{A}31}^{\circ}=1/3,\quad M_{\text{A}41}^{\circ}=1/2,\quad M_{\text{A}5}^{\circ}=1/5,
\]
also the region with ferromagnetic ordering is equal to unity,
\[
M_{\text{F}1}^{\circ}=1.
\]

In the presence of an external magnetic field at the junctions of
configurations of the MPDGS, the residual entropy and residual magnetization
have a wide variety of values, demonstrating both the absence and
the presence of frustrations in the corresponding regions of the phase
diagram.

In the ground state, at the boundaries of configurations $C_{\text{A}2}$–$C_{\text{A}4}$
and $C_{\text{A}2}$–$C_{\text{A}6}$, the residual specific entropy
and residual magnetization are zero,
\begin{equation}
S_{\text{A}2-\text{A}4}^{\circ}=0,\quad M_{\text{A}2-\text{A}4}^{\circ}=0,\label{eq:S0:M0:A2:A4}
\end{equation}
\begin{equation}
S_{\text{A}2-\text{A}6}^{\circ}=0,\quad M_{\text{A}2-\text{A}6}^{\circ}=0\label{eq:S0:M0:A2:A6}
\end{equation}
(green dashed lines on the MPDGS, Figs.~\ref{fig:N3h:PD:J3:J2:m}–
\ref{fig:N3h:PD:H:J3:pp}). Recall that on the MPDGS the lines with
zero residual entropy are marked with a dashed line.

At the boundaries $C_{\text{A}2}$–$C_{\text{A}5}$, $C_{\text{F}1}$–$C_{\text{A}2}$,
$C_{\text{F}1}$–$C_{\text{A}4}$, $C_{\text{F}1}$–$C_{\text{A}31}$,
the residual entropy and residual magnetization are, respectively,
\begin{equation}
S_{\text{A}2-\text{A}5}^{\circ}=0,\quad M_{\text{A}2-\text{A}5}^{\circ}=1/10,\label{eq:S0:M0:A2:A5}
\end{equation}
\begin{equation}
S_{\text{F}1-\text{A}2}^{\circ}=0,\quad M_{\text{F}1-\text{A}2}^{\circ}=1/2,\label{eq:S0:M0:F1:A2}
\end{equation}
\begin{equation}
S_{\text{F}1-\text{A}4}^{\circ}=0,\quad M_{\text{F}1-\text{A}4}^{\circ}=1/2,\label{eq:S0:M0:F1:A4}
\end{equation}
\begin{equation}
S_{\text{F}1-\text{A}31}^{\circ}=0,\quad M_{\text{F}1-\text{A}31}^{\circ}=2/3\label{eq:S0:M0:F1:A31}
\end{equation}
(blue dashed lines on the MPDGS, Figs.~\ref{fig:N3h:PD:J3:J2:m}–
\ref{fig:N3h:PD:H:J3:pp}).

Next, we consider the remaining cases where the residual entropy of
the spin system is greater than zero. (Such states on the MPDGS are
marked with red solid lines, and phase points are marked with red
round dots.)

1) At the boundary of the regions of spin configurations $C_{\text{A}2}$–$C_{\text{A}31}$
and at the triple point $C_{\text{F}1}$–$C_{\text{A}2}$–$C_{\text{A}31}$
(see, for example, Fig.~\ref{fig:N3h:PD:J3:J2:m}b), the residual
entropy is
\begin{equation}
S^{\circ}=\ln\left[\frac{1}{3}\left(\vartheta_{1}+\frac{3}{\vartheta_{1}}\right)\right]\approx0.281\,199\,6,\label{eq:N3h:S0:281:H}
\end{equation}
where the value of $\vartheta_{1}$ is the same as in the case (\ref{eq:N3h:S0:281:T1}),
and the residual magnetization is
\[
M^{\circ}=\frac{1}{3}\left(1+\zeta_{1}+\frac{4}{\mathcal{D}_{1}\zeta_{1}}\right)\approx0.177\,008\,8,
\]
where
\[
\zeta_{1}=\sqrt[3]{4\frac{\mathcal{D}_{1}+3\sqrt{-3\mathcal{D}_{1}}}{\mathcal{D}_{1}^{2}}}.
\]
In this case, the expression for the residual magnetization is the
principal solution to the equation
\[
-\mathcal{D}_{1}(y^{3}-y^{2})+3^{2}y-1=0,
\]
where $\mathcal{D}_{1}=-23$ is the discriminant of the equation (\ref{eq:N3h:S0:281:E0}).

Also, at the boundary of spin configurations $C_{\text{F}1}$–$C_{\text{A}5}$
(see, for example, Fig.~\ref{fig:N3h:PD:J3:J2:p}b), the residual
entropy is (\ref{eq:N3h:S0:281:H}), and the residual magnetization
is
\[
M^{\circ}\approx0.504\,925\,7.
\]

At the triple point $C_{\text{A}2}$–$C_{\text{A}41}$–$C_{\text{A}5}$
(see Fig.~\ref{fig:N3h:PD:J3:J2:m}d) the residual entropy is (\ref{eq:N3h:S0:281:H}),
and the residual magnetization is
\[
M^{\circ}\approx0.280\,999\,7.
\]

At the intersections of spin configurations at points $C_{\text{A}4}$–$C_{\text{A}5}$–$C_{\text{A}6}$
and $C_{\text{A}4}$–$C_{\text{A}5}$–($C_{\text{A}2}$+$C_{\text{A}6}$),
the residual entropy is defined in (\ref{eq:N3h:S0:281:H}), and the
residual magnetization is zero,
\[
M^{\circ}=0.
\]

2) At the triple point $C_{\text{F}1}$–$C_{\text{A}31}$–$C_{\text{A}41}$
(see Fig.~\ref{fig:N3h:PD:J3:J2:m}f), the residual entropy is
\begin{equation}
S^{\circ}=\ln\left[\frac{1}{3}\left(1+\vartheta_{2}+\frac{1}{\vartheta_{2}}\right)\right]\approx0.382\,245\,1,\label{eq:N3h:S0:382:H}
\end{equation}
where $\vartheta_{2}$ is defined in (\ref{eq:N3h:S0:382:T1}), and
the residual magnetization is
\[
M^{\circ}=\frac{1}{3}\left(1+\vartheta_{2}+\frac{4}{\mathcal{D}_{2}\vartheta_{2}}\right)\approx0.611\,492\,0,
\]
where
\[
\zeta_{2}=\sqrt[3]{4\frac{-\mathcal{D}_{2}+3\sqrt{-3\mathcal{D}_{2}}}{\mathcal{D}_{2}^{2}}}.
\]
In this case, the expression for the residual magnetization is the
principal solution to the equation
\[
-\mathcal{D}_{2}(y^{3}-y^{2})+3^{2}y-1=0,
\]
where $\mathcal{D}_{2}=-31$ is the discriminant of the equation (\ref{eq:N3h:S0:382:E0}).

Also at the triple point $C_{\text{A}2}$–$C_{\text{A}31}$–$C_{\text{A}41}$
(see Fig.~\ref{fig:N3h:PD:J3:J2:m}d) the residual entropy is (\ref{eq:N3h:S0:382:H}),
and the residual magnetization is
\[
M^{\circ}\approx0.273\,040\,6.
\]

3) At the intersections of spin configurations at the points of the
phase diagram $C_{\text{F}1}$–$C_{\text{A}2}$–$C_{\text{A}31}$–$C_{\text{A}41}$
and $C_{\text{F}1}$–$C_{\text{A}2}$–$C_{\text{A}31}$(–$C_{\text{A}41}$)
($J_{3}=0$) (see Fig.~\ref{fig:N3h:PD:J3:J2:m}e), the residual
entropy is equal to the natural logarithm of the golden ratio,
\[
S^{\circ}=\ln\frac{1+\sqrt{5}}{2}\approx0.481\,211\,8,
\]
as in (\ref{eq:N3h:S0:481}). In this case, the residual magnetization
is equal to
\[
M^{\circ}=\frac{1}{\sqrt{\mathcal{D}_{3}}}\approx0.447\,213\,6,
\]
where the expression for the residual magnetization is the principal
solution of the equation
\[
\mathcal{D}_{3}y^{2}-1=0,
\]
where $\mathcal{D}_{3}=5$ is the discriminant of the equation (\ref{eq:N3h:S0:481:E0}).

4) At the boundaries of spin configurations $C_{\text{A}31}$–$C_{\text{A}4}$,
and at the triple point $C_{\text{F}1}$–$C_{\text{A}31}$–$C_{\text{A}4}$
(see Figs.~\ref{fig:N3h:PD:J3:J2:m}b or \ref{fig:N3h:PD:J3:J2:p}b),
the residual entropy is
\begin{equation}
S^{\circ}=\ln\left(-\psi_{4}+\frac{1}{2}\sqrt{-\frac{1}{\psi_{4}}-4\psi_{4}^{2}}\right)\approx0.199\,460\,6,\label{eq:N3h:S0:199:H}
\end{equation}
where
\[
\psi_{4}=\frac{1}{2}\sqrt{\frac{\vartheta_{4}}{3}-\frac{4}{\vartheta_{4}}},\quad\vartheta_{4}=\sqrt[3]{\frac{3^{3}+3\sqrt{-3\mathcal{D}_{4}}}{2}}.
\]
That the sublogarithmic expression is the principal solution of the
equation
\[
x^{4}-x-1=0,
\]
its discriminant is
\[
\mathcal{D}_{4}=-(2^{8}+3^{3})=-283.
\]
In this case, the residual magnetization is equal to
\[
M^{\circ}=-\xi_{4}+\sqrt{-\frac{3^{2}}{\mathcal{D}_{4}}-\xi_{4}^{2}-\frac{2}{\mathcal{D}_{4}\xi_{4}}}\approx0.159\,319\,6,
\]
where
\[
\xi_{4}=\frac{1}{2}\sqrt{\frac{\zeta_{4}}{3}-\frac{2^{10}}{\mathcal{D}_{4}^{2}\zeta_{4}}-3\frac{2^{2}}{\mathcal{D}_{4}}},
\]
\[
\zeta_{4}=\sqrt[3]{-2^{11}3\frac{3^{2}+\sqrt{-3\mathcal{D}_{4}}}{\mathcal{D}_{4}^{3}}}.
\]
Here, the expression for the residual magnetization is the principal
solution of the equation
\[
-\mathcal{D}_{4}y^{4}-2(3^{2}y^{2}-2^{2}y)-1=0.
\]

Also, at the boundary of spin configurations $C_{\text{A}31}$–$C_{\text{A}41}$
(see Fig.~\ref{fig:N3h:PD:J3:J2:m}d), the residual entropy is also
equal to (\ref{eq:N3h:S0:199:H}), and the residual magnetization
is
\[
M^{\circ}\approx0.420\,340\,2.
\]

5) At the triple point $C_{\text{F}1}$–$C_{\text{A}2}$–$C_{\text{A}41}$
(see Fig.~\ref{fig:N3h:PD:J3:J2:m}f) the residual entropy is
\begin{equation}
S^{\circ}=\ln\left(\frac{1}{4}-\psi_{5}+\frac{1}{2}\sqrt{\frac{11}{4}+\frac{3}{8\psi_{5}}-4\psi_{5}^{2}}\right)\approx0.414\,012\,7,\label{eq:N3h:S0:414:H}
\end{equation}
where
\[
\psi_{5}=\frac{1}{2}\sqrt{\frac{11}{12}+\frac{1}{3}\left(\vartheta_{5}-\frac{8}{\vartheta_{5}}\right)},
\]
\[
\vartheta_{5}=\sqrt[3]{\frac{-2-3^{4}+3\sqrt{-3\mathcal{D}_{5}}}{2}}.
\]
The sublogarithmic expression is the principal solution of the equation
\[
x^{4}-x^{3}-x^{2}+x-1=0,
\]
its discriminant is
\[
\mathcal{D}_{5}=-[3(2+3)^{2}+2^{8}]=-331.
\]
In this case, the residual magnetization is equal to
\[
M^{\circ}\approx0.526\,524\,3.
\]

6) At the boundary of spin configurations $C_{\text{F}1}$–$C_{\text{A}4}$
($J_{3}=0$) or at the triple point $C_{\text{F}1}$–$C_{\text{A}4}$–$C_{\text{A}5}$
(see Fig.~\ref{fig:N3h:PD:J3:J2:p}b), the residual entropy is
\begin{equation}
S^{\circ}=\ln\left(\frac{1}{4}-\psi_{6}+\frac{1}{2}\sqrt{\frac{3}{4}-\frac{1}{8\psi_{6}}-4\psi_{6}^{2}}\right)\approx0.322\,284\,6,\label{eq:N3h:S0:322:H}
\end{equation}
where
\[
\psi_{6}=\frac{1}{2}\sqrt{\frac{1}{4}+\frac{\vartheta_{6}}{3}-\frac{4}{\vartheta_{6}}},\quad\vartheta_{6}=\sqrt[3]{-\frac{3^{3}-3\sqrt{-3\mathcal{D}_{6}}}{2}}.
\]
The sublogarithmic expression is the principal solution of the equation
\[
x^{4}-x^{3}-1=0,
\]
its discriminant is equal to
\[
\mathcal{D}_{6}=-(2^{8}+3^{3})=-283.
\]
In this case, the residual magnetization is equal to
\[
M^{\circ}=-\xi_{6}+\sqrt{-\frac{3^{2}}{\mathcal{D}_{6}}-\xi_{6}^{2}+\frac{2}{\mathcal{D}_{6}\xi_{6}}}\approx0.396\,650\,6,
\]
where
\[
\xi_{6}=\frac{1}{2}\sqrt{\frac{\zeta_{6}}{3}-\frac{2^{10}}{\mathcal{D}_{6}^{2}\zeta_{6}}-3\frac{2^{2}}{\mathcal{D}_{6}}},
\]
\[
\zeta_{6}=\sqrt[3]{-2^{11}3\frac{3^{2}+\sqrt{-3\mathcal{D}_{6}}}{\mathcal{D}_{6}^{3}}}.
\]
Moreover, the expression for the residual magnetization is the principal
solution of the following equation
\[
-\mathcal{D}_{7}y^{4}-2(3^{2}y^{2}+2^{2}y)-1=0.
\]

Also, at the boundary of spin configurations $C_{\text{F}1}$–$C_{\text{A}41}$
(see Fig.~\ref{fig:N3h:PD:J3:J2:m}f), the residual entropy is (\ref{eq:N3h:S0:322:H}),
and the residual magnetization is
\[
M^{\circ}\approx0.698\,325\,3.
\]

7) At the boundary of spin configurations $C_{\text{A}2}$–$C_{\text{A}41}$
(see Fig.~\ref{fig:N3h:PD:J3:J2:m}d) the residual entropy is equal
to the natural logarithm of the golden ratio square root,
\[
S^{\circ}=\ln\sqrt{\frac{1+\sqrt{5}}{2}}\approx0.240\,605\,9,
\]
where the sublogarithmic expression is the principal solution of the
equation
\[
x^{4}-x^{2}-1=0.
\]
And the residual magnetization is equal to
\[
M^{\circ}\approx0.276\,393\,2.
\]

8) At the boundary of spin configurations $C_{\text{A}41}$–$C_{\text{A}5}$
(see Fig.~\ref{fig:N3h:PD:J3:J2:m}d), the residual entropy is equal
to
\begin{equation}
S^{\circ}=\ln\psi_{8}\approx0.154\,696\,8,\label{eq:N3h:S0:154:H}
\end{equation}
where the sublogarithmic expression is the principal solution of the
equation
\[
x^{5}-x-1=0.
\]
In this case, the residual magnetization is equal to
\[
M^{\circ}\approx0.344\,868\,5.
\]

Also at the boundary of spin configurations $C_{\text{A}4}$–$C_{\text{A}5}$
and at the triple point $C_{\text{A}2}$–$C_{\text{A}4}$–$C_{\text{A}5}$
(see Fig.~\ref{fig:N3h:PD:J3:J2:m}b), the residual entropy is equal
to (\ref{eq:N3h:S0:154:H}), and the residual magnetization is
\[
M^{\circ}\approx0.103\,421\,0.
\]

9) At the boundary of spin configurations ($C_{\text{A}31}$–)$C_{\text{A}4}$–$C_{\text{A}41}$(–$C_{\text{A}5}$),
and at the point $C_{\text{A}31}$–$C_{\text{A}4}$–$C_{\text{A}41}$–$C_{\text{A}5}$
(see Fig.~\ref{fig:N3h:PD:J3:J2:m}d), the residual entropy is
\[
S^{\circ}=\ln\psi_{9}\approx0.354\,382\,0,
\]
where the sublogarithmic expression is the principal solution of the
equation
\[
x^{5}-x^{2}-2x-1=0.
\]
The residual magnetization is
\[
M^{\circ}\approx0.261\,462\,5.
\]

10) At the boundaries of spin configurations $C_{\text{A}5}$–$C_{\text{A}6}$
or $C_{\text{A}2}$+$C_{\text{A}6}$–$C_{\text{A}5}$, and at the
triple point $C_{\text{A}2}$–$C_{\text{A}5}$–$C_{\text{A}6}$ (see
Figs.~\ref{fig:N3h:PD:J3:J2:m}b or \ref{fig:N3h:PD:H:J3:mm}e),
the residual entropy is
\[
S^{\circ}=\ln\psi_{10}\approx0.126\,389\,6,
\]
where the sublogarithmic expression is the principal solution of the
equation
\[
x^{6}-x-1=0,
\]
and the residual magnetization is
\[
M^{\circ}\approx0.097\,204\,1.
\]

11) At the boundaries of spin configurations $C_{\text{F}1}$–$C_{\text{A}6}$
and $C_{\text{F}1}$–$C_{\text{A}2}$+$C_{\text{A}6}$, including
the triple point $C_{\text{F}1}$–$C_{\text{A}2}$–$C_{\text{A}6}$
(see Figs.~\ref{fig:N3h:PD:J3:J2:p}b or \ref{fig:N3h:PD:H:J3:pp}b),
the residual entropy is
\[
S^{\circ}=\ln\psi_{11}\approx0.250\,913\,6,
\]
where the sublogarithmic expression is the principal solution of the
equation
\[
x^{6}-x^{5}-1=0,
\]
and the residual magnetization is
\[
M^{\circ}\approx0.368\,841\,2.
\]

12) At the triple point $C_{\text{F}1}$–$C_{\text{A}5}$–$C_{\text{A}6}$
(see Fig.~\ref{fig:N3h:PD:J3:J2:p}b), the residual entropy is
\[
S^{\circ}=\ln\psi_{12}\approx0.350\,398\,2,
\]
where the sublogarithmic expression is the principal solution of the
equation
\[
x^{6}-x^{5}-x-1=0,
\]
and the residual magnetization is
\[
M^{\circ}\approx0.380\,916\,8.
\]

13) At the triple point $C_{\text{A}2}$–$C_{\text{A}31}$–$C_{\text{A}4}$
(see Fig.~\ref{fig:N3h:PD:J3:J2:m}b), the residual entropy is
\[
S^{\circ}=\ln\psi_{13}\approx0.337\,377\,8,
\]
where the sublogarithmic expression is the principal solution of the
equation
\[
x^{6}-x^{4}-x^{3}-x^{2}+1=0,
\]
and the residual magnetization is
\[
M^{\circ}\approx0.154\,405\,9.
\]

14) At the points of the phase diagram $C_{\text{A}2}$–$C_{\text{A}31}$–$C_{\text{A}4}$–$C_{\text{A}41}$–$C_{\text{A}5}$
or $C_{\text{A}2}$–$C_{\text{A}4}$–$C_{\text{A}41}$(–$C_{\text{A}31}$–$C_{\text{A}5}$)
(see Figs.~\ref{fig:N3h:PD:J3:J2:m}c, \ref{fig:N3h:PD:H:J2:mm}c
or \ref{fig:N3h:PD:H:J3:mm}c), the residual entropy is
\[
S^{\circ}=\ln\psi_{14}\approx0.446\,997\,7,
\]
where the sublogarithmic expression is the principal solution of the
equation
\[
x^{6}-x^{5}-x^{3}-x^{2}+1=0,
\]
the residual magnetization is
\[
M^{\circ}\approx0.238\,982\,9.
\]

\begin{figure}
\centering\includegraphics{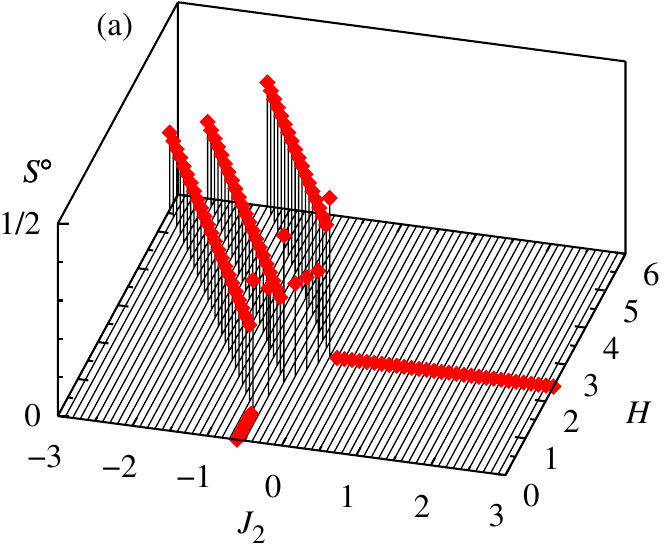}\quad{}\includegraphics{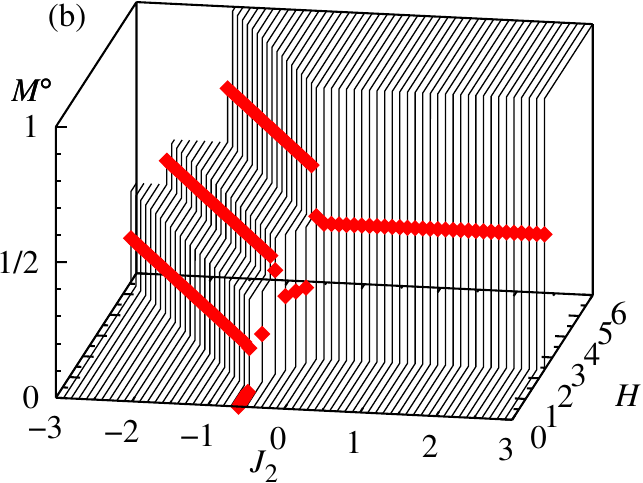}
\caption{Entropy (a) and magnetization (b) of the ground state of the Ising
chain in an external magnetic field, taking into account the interaction
of spins at the sites of the first, second, and third neighbors with
antiferromagnetic interaction of nearest neighbors ($J_{1}=-1$) and
antiferromagnetic interaction of third neighbors $(J_{3}=-1/5$).
Red rhombic dots indicate the values of the functions at the boundaries
of the spin configurations}
\label{fig:N3h:S0:M0:H:J2:3b}
\end{figure}

\begin{figure}
\centering\includegraphics{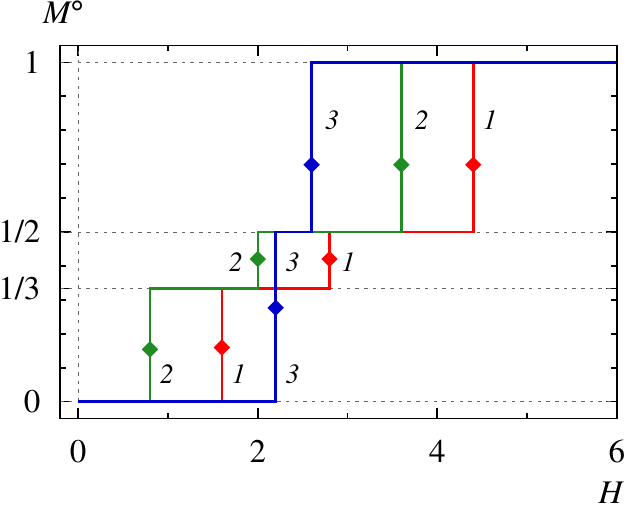}
\caption{Magnetization of the ground state of the Ising chain in an external
magnetic field, taking into account the exchange interaction of spins
at the sites of the first, second, and third neighbors with antiferro-/antiferro-/antiferromagnetic
exchange interactions, where $J_{1}=-1$, $J_{3}=-1/5$, $J_{2}=-1$
(red line~1), $J_{2}=-3/5$ (green line~2), and $J_{2}=-1/10$ (blue
line~3). Rhombic dots mark frustration magnetization points}
\label{fig:N3h:S0:M0:H:J2:3b:}
\end{figure}

\begin{figure}
\centering\includegraphics{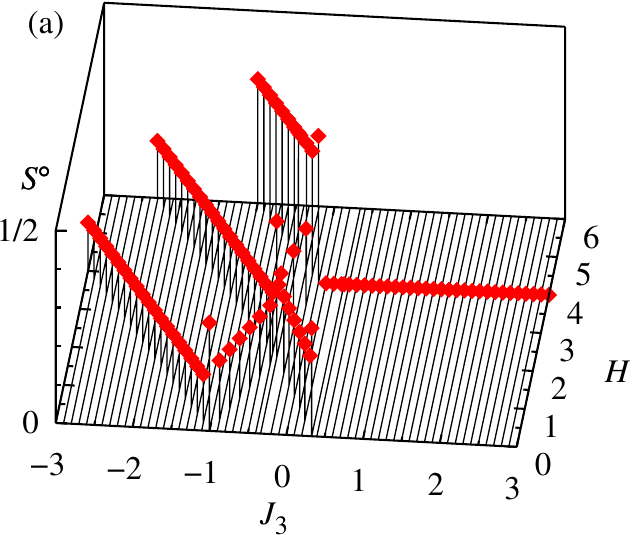}\quad{}\includegraphics{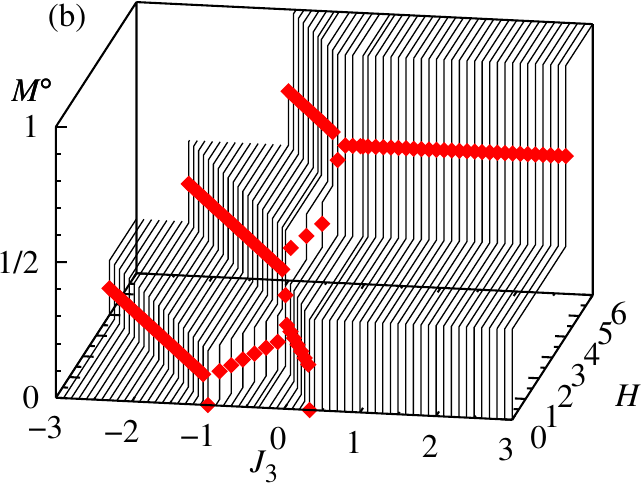}
\caption{Entropy (a) and magnetization (b) of the ground state of the Ising
chain in an external magnetic field, taking into account the interaction
of spins at the sites of the first, second, and third neighbors with
antiferromagnetic interaction of nearest neighbors ($J_{1}=-1$) and
antiferromagnetic interaction of the second neighbors ($J_{2}=-1$).
Red rhombic dots indicate the values of the functions at the boundaries
of the spin configurations}
\label{fig:N3h:S0:M0:H:J3:7e}
\end{figure}

\begin{figure}
\centering\includegraphics{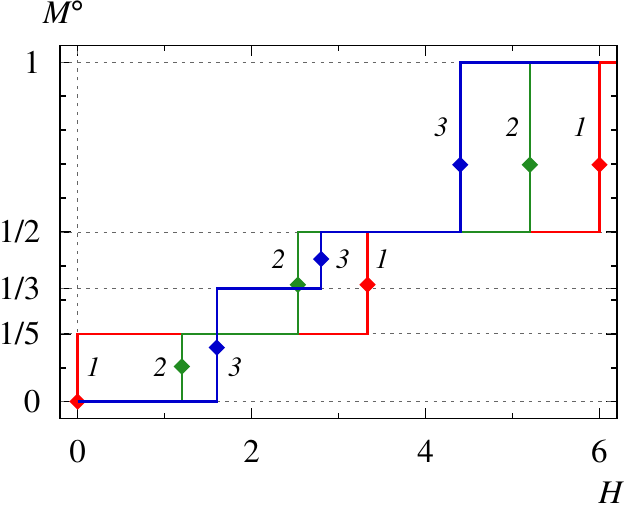}
\caption{Magnetization of the ground state of the Ising chain in an external
magnetic field, taking into account the exchange interaction of spins
at the sites of the first, second, and third neighbors with antiferro/antiferro/antiferromagnetic
exchange interactions, where $J_{1}=-1$, $J_{2}=-1$, $J_{3}=-1$
(red line~1), $J_{3}=-3/5$ (green line~2), and $J_{3}=-1/5$ (blue
line~3). Rhombic dots mark frustration magnetization points}
\label{fig:N3h:S0:M0:J3:J2:7e:}
\end{figure}

As an illustration, consider several examples of the behavior of the
residual entropy and residual magnetization in Figs.~\ref{fig:N3h:S0:M0:H:J2:3b}
and \ref{fig:N3h:S0:M0:H:J3:7e}, which correspond to the phase diagrams
shown in Figs.~\ref{fig:N3h:PD:H:J2:mm}b and \ref{fig:N3h:PD:H:J3:mm}e.

In Figs.~\ref{fig:N3h:S0:M0:H:J2:3b:} and \ref{fig:N3h:S0:M0:J3:J2:7e:}
the field dependences of residual magnetization are shown for several
ratios of the values of exchange interactions presented in Figs.~\ref{fig:N3h:S0:M0:H:J2:3b}
and \ref{fig:N3h:S0:M0:H:J3:7e}.

In this case, note that in Fig.~\ref{fig:N3h:S0:M0:J3:J2:7e:} (line~1)
in the case of antiferro-antiferro-antiferromagnetic exchange interaction
in the spin chain, a non-trivial behavior of the spin system arises
when a non-zero magnetization in the ground state is retained at $H=0$.

It should be noted that at zero temperature ($T=0$) in the presence
of an external magnetic field ($H>0$), the other thermodynamic functions
of the spin system, such as the residual heat capacity (\ref{eq:CV}),
for any model parameters, is always equal to zero,
\[
C^{\circ}=0.
\]
The residual magnetic susceptibility (\ref{eq:CHI}) of the system
beyond the boundaries of spin configurations and at the boundaries
$C_{\text{A}2}$–$C_{\text{A}4}$ and $C_{\text{A}2}$–$C_{\text{A}6}$
(marked with green dashed lines on the MPDGS) is zero,
\[
\chi^{\circ}=0,
\]
at the boundaries of spin configurations $C_{\text{A}2}$–$C_{\text{A}5}$,
$C_{\text{F}1}$–$C_{\text{A}2}$, $C_{\text{F}1}$–$C_{\text{A}4}$,
$C_{\text{F}1}$–$C_{\text{A}31}$ (marked with blue dashed lines
on the MPDGS) has values in the interval
\[
0<\chi^{\circ}<\infty,
\]
and in the case of frustrated states, the residual magnetic susceptibility
is equal to infinity,
\[
\chi^{\circ}=\infty.
\]
This situation is of a special interest, which should be considered
in the next following paper.

In the end, it should be noted that the Ising paramagnet in the absence
of all exchange interactions of spins between neighbors ($J_{i}=0$)
in an external magnetic field ($H>0$) in the ground state ($T=0$)
is characterized by only one configuration $C_{\text{F}1}$ (\ref{eq:C:F1}),
in which the chain spins are oriented along the direction of an external
magnetic field. Therefore, the residual entropy and residual magnetization
of the system are respectively equal to
\[
S^{\circ}=0,\quad M^{\circ}=1.
\]
There are no relevant frustrated states in the considered range of
model parameters.

\section{Conclusions}

In this paper, the precise analytical expressions for the entropy,
heat capacity, magnetization, and magnetic susceptibility of the one-dimensional
Ising model in an external magnetic field, taking into account the
exchange interactions of atomic spins at the sites of the first, second,
and third neighbors are obtained by the Kramers--Wannier transfer-matrix
method. The analysis of the configuration features of the ground state,
the description of the boundaries of the transitions of spin configurations
and the frustrating properties of the spin system under study are
carried out; a complete magnetic phase diagram of the ground state
model is constructed.

The criteria are formulated and the relations of the model parameters
at which magnetic frustrations occur in the considered one-dimensional
spin systems are determined. It was found out that frustrations are
caused by competition between the energies of the exchange interactions
of spins and an external magnetic field. Thus, it is shown that in
the frustration regime, the system undergoes a rearrangement of the
structure of the magnetic ordering in the ground state, which begins
to include a set of spin configurations comparable to the size of
the system, including those without any translational invariance.

The behavior of entropy, magnetization, heat capacity, and magnetic
susceptibility in the ground state of the system is analyzed.

A cardinal difference in the behavior of the entropy in the ground
state of the magnetic system in the frustration region and beyond
it is shown. It is determined that the most important attribute of
the existence of magnetic frustrations in the system is the non-zero
value of the zero-temperature entropy in this regime, and that this
property does not contradict the third law of thermodynamics.

The values of entropy and magnetization for all configurations of
the ground state of the spin system depending on the values of the
model parameters are calculated. The features of the behavior of the
heat capacity and magnetic susceptibility of the system at zero temperature
are considered.

The paper also compares the behavior of the entropy, magnetization,
and magnetic susceptibility of the system at zero temperature with
the magnetic phase diagram of the ground state.

It is found that the entropy, magnetization, and magnetic susceptibility
of the ground state exhibit several types of behavior at the boundaries
of spin configurations, depending on the presence or absence of frustrations
in the spin system.

It is also noted that at a certain ratio of the antiferro-antiferro-antiferromagnetic
parameters of the exchange interactions of the model, the spin system
can have a non-zero magnetization in the absence of a field and at
zero temperature.

As a special example, it is demonstrated that an Ising paramagnetic,
which in the absence of an external magnetic field is an absolutely
frustrated system, since its entropy is nonzero and does not depend
on temperature.

Thus, the proposed analysis scheme allows us to consider a wide range
of phenomena in one-dimensional (or quasi-one-dimensional) magnetic
systems with frustrations and describe their relationship with the
peculiar features of thermodynamic functions. The mathematical apparatus
developed in the present paper makes it possible to solve similar
problems in more complicated models of statistical physics, in particular,
in multicomponent spin models with discrete symmetry and arbitrary
spin value.

\section*{Acknowledgment}

The research was carried out within the state assignment of Ministry of Science and Higher Education of the Russian Federation (theme ``Quantum'' No. 122021000038-7).

\end{document}